\documentclass[%reprint,
superscriptaddress,
nofootinbib,
 twocolumn,
 amsmath,amssymb,
 aps,
]{revtex4-1}

\usepackage[utf8]{inputenc}
\usepackage{graphicx}
\usepackage{xcolor}
\usepackage{hyperref}
\usepackage{siunitx}
\usepackage[normalem]{ulem}
\newcommand{\net}{\ensuremath{\mathrm{net}}}
\newcommand{\data}{\ensuremath{s}}
\newcommand{\vdata}{\ensuremath{\vec{\data}}}
\newcommand{\noise}{\ensuremath{n}}
\newcommand{\vnoise}{\ensuremath{\vec{\noise}}}
\newcommand{\signal}{\ensuremath{h}}
\newcommand{\vsignal}{\ensuremath{\vec{\signal}}}

\newcommand{\vcovmat}{\ensuremath{\vec{C}}}

\newcommand{\numdet}{\ensuremath{K}}
\newcommand{\ip}[2]{\left< #1, #2 \right>}
\newcommand{\psd}{\ensuremath{S_n}}
\newcommand{\lmn}{{\ell m n}}
\newcommand{\mdot}{\mathrm{M}_{\odot}}
\newcommand{\signalinj}{Signal}
\newcommand{\controlinj}{Control}
\newcommand{\bayesfac}{\ensuremath{\mathcal{B}}} % bayes factor variable
\newcommand{\eventbftime}{\ensuremath{t_{\rm ref} + 6\,\mathrm{ms}}} % GW190521 max BF time
\newcommand{\eventbf}{{\ensuremath{56 \pm 1 }}} % GW190521 330 BF
\newcommand{\approxebf}{\ensuremath{56}} % approximate GW190521 BF
\newcommand{\expectedFA}{8.9} % the number of low q injections expected to have BF > 190521
\newcommand{\measuredFA}{10} % the number of low q injections with BF > 190521
\newcommand{\measuredFAallt}{16}
\newcommand{\numhiqinj}{45}
\newcommand{\measuredTA}{22}
\newcommand{\measuredTAallt}{23}

\usepackage{array}
\newcolumntype{P}[1]{>{\centering\arraybackslash}p{#1}}

\begin{document}

\title{Estimating False Alarm Rates of Sub-Dominant Quasi-normal Modes in GW190521}

\author{Collin D. Capano}
\email{corresponding author}
\affiliation{Department of Physics, Syracuse University, Syracuse, NY 13244, USA}
\affiliation{Department of Physics, University of Massachusetts, Dartmouth, MA 02747, USA}
\affiliation{Max-Planck-Institut f{\"u}r Gravitationsphysik (Albert-Einstein-Institut), Callinstra{\ss}e 38, 30167 Hannover, Germany}
\affiliation{Leibniz Universit{\"a}t Hannover, 30167 Hannover, Germany}
\author{Jahed Abedi}
\affiliation{Department of Mathematics and Physics, University of Stavanger, NO-4036 Stavanger, Norway}
\affiliation{Max-Planck-Institut f{\"u}r Gravitationsphysik (Albert-Einstein-Institut), Callinstra{\ss}e 38, 30167 Hannover, Germany}
\affiliation{Leibniz Universit{\"a}t Hannover, 30167 Hannover, Germany}
\author{Shilpa Kastha}
\affiliation{Niels Bohr International Academy, Niels Bohr Institute, Blegdamsvej 17, 2100 Copenhagen, Denmark}
\affiliation{Max-Planck-Institut f{\"u}r Gravitationsphysik (Albert-Einstein-Institut), Callinstra{\ss}e 38, 30167 Hannover, Germany}
\affiliation{Leibniz Universit{\"a}t Hannover, 30167 Hannover, Germany}
\author{Alexander H. Nitz}
\affiliation{Department of Physics, Syracuse University, Syracuse, NY 13244, USA}
\affiliation{Max-Planck-Institut f{\"u}r Gravitationsphysik (Albert-Einstein-Institut), Callinstra{\ss}e 38, 30167 Hannover, Germany}
\affiliation{Leibniz Universit{\"a}t Hannover, 30167 Hannover, Germany}
\author{Julian Westerweck}
\affiliation{Max-Planck-Institut f{\"u}r Gravitationsphysik (Albert-Einstein-Institut), Callinstra{\ss}e 38, 30167 Hannover, Germany}
\affiliation{Leibniz Universit{\"a}t Hannover, 30167 Hannover, Germany}
\author{Yi-Fan Wang}
\affiliation{Max-Planck-Institut f{\"u}r Gravitationsphysik (Albert-Einstein-Institut), Am M{\"u}hlenberg 1, D-14476 Potsdam, Germany}
\affiliation{Max-Planck-Institut f{\"u}r Gravitationsphysik (Albert-Einstein-Institut), Callinstra{\ss}e 38, 30167 Hannover, Germany}
\affiliation{Leibniz Universit{\"a}t Hannover, 30167 Hannover, Germany}
\author{Miriam Cabero}
\affiliation{Department of Physics and Astronomy, The University of British Columbia, Vancouver, BC V6T 1Z4, Canada}
\author{Alex B. Nielsen}
\affiliation{Department of Mathematics and Physics, University of Stavanger, NO-4036 Stavanger, Norway}
\author{Badri Krishnan}
\affiliation{Institute for Mathematics, Astrophysics and Particle Physics, Radboud University, Heyendaalseweg 135, 6525 AJ Nijmegen, The Netherlands}
\affiliation{Max-Planck-Institut f{\"u}r Gravitationsphysik (Albert-Einstein-Institut), Callinstra{\ss}e 38, 30167 Hannover, Germany}
\affiliation{Leibniz Universit{\"a}t Hannover, 30167 Hannover, Germany}

\begin{abstract}

  A major aim of gravitational wave astronomy is to test
  observationally the Kerr nature of black holes.  The strongest
  such test, with minimal additional assumptions, is provided by 
  observations of multiple ringdown modes,
  also known as black hole spectroscopy. For the gravitational wave
  merger event GW190521, we have previously claimed the detection
  of two ringdown modes emitted by the remnant black hole.
  In this paper we
  provide further evidence for the detection of multiple ringdown
  modes from this event.  We analyse the recovery of simulated
  gravitational wave signals designed to replicate the ringdown
  properties of GW190521. We quantify how often our detection
  statistic reports strong evidence for a sub-dominant $(\ell,m,n)=(3,3,0)$
  ringdown mode, even when no such mode is present in the simulated signal.
  We find this only occurs with a probability $\sim 0.02$, which is
  consistent with a Bayes factor of $\eventbf{}$ (1$\sigma$ uncertainty) found for GW190521. 
  We also quantify our agnostic analysis of GW190521, in which
  no relationship is assumed between ringdown modes, and find
  that only 1 in 250 simulated signals without a $(3,3,0)$
  mode yields a result as significant as GW190521. Conversely, we verify
  that when simulated signals do have an observable $(3,3,0)$ mode
  they consistently yield a strong evidence and significant agnostic results.
  We also find that constraints on deviations from the $(3,3,0)$ mode on
  GW190521-like signals with a $(3,3,0)$ mode are consistent with what was
  obtained from our previous analysis of GW190521.
  Our results support our previous conclusion that the gravitational wave signal from
  GW190521 contains an observable sub-dominant $(\ell,m,n)=(3,3,0)$ mode.

\end{abstract}

\maketitle

\section{Introduction}
%***************************************************

Einstein's theory of general relativity (GR) predicts that black holes are stable to
perturbations \cite{Andersson:2019dwi}. A distorted black hole should settle down to a
stationary Kerr state through the emission of gravitational waves
\cite{Whiting:1988vc}. This applies to the remnant black hole formed 
in a binary black hole merger event, which is highly distorted on 
formation, but is expected to eventually settle down to a Kerr black 
hole due to the emission of gravitational waves. According to theory, the gravitational 
waveform in the late stages of a merger event consists of a spectrum of quasi-normal modes with a rich structure of
different fundamental modes and overtones \cite{Berti:2009kk}. Several numerical studies confirm this expectation that the remnant black hole settles down to equilibrium by emitting quasi-normal modes; see e.g. \cite{Mourier:2020mwa,Gupta:2018znn,Campanelli:2008dv,Owen:2009sb}.

The spectrum consists of a set of complex frequencies (determined by the black hole mass and spin) labeled by three integers $(\ell,m,n)$, with $\ell\geq 2$, $-\ell\leq m \leq \ell$, and $n\geq 0$.  Modes with $n\geq 1$ are known as ``overtones".  
Using black hole spectroscopy, the observation of more than one such
ringdown mode can be used to determine if the black hole is consistent
with GR \cite{Dreyer:2003bv,Kamaretsos:2011um}.  A
clear and unambiguous determination of multiple ringdown modes
provides one of the strongest tests of the Kerr nature of black holes
in our universe and a possible route to discover new physics beyond
standard general relativity.  See \cite{Gossan:2011ha} for a general Bayesian formulation for implementing such tests of general relativity.  

Quasi-normal modes for a Schwarzschild black hole were first
identified by Vishveshwara
\cite{Vishveshwara:1970zz,Vishveshwara:1970cc}, and further studied
within black hole perturbation theory by Chandrasekhar and Detweiler
\cite{Chandrasekhar:1975zza}. They were suggested to be the gravitational waves emitted by a perturbed black hole in \cite{Press:1971wr}.  
There remain several outstanding
theoretical questions regarding black hole quasi-normal modes which
are also important for observational studies.  The first is the
question of the start time for the ringdown. When the remnant black
hole is formed, it is initially highly distorted away from a Kerr
black hole. The black hole loses these distortions over time, and at
some point it can be considered to be a linear perturbation of a Kerr
black hole.  There is some evidence that we can go closer to the merger with just the usual quasi-normal modes defined within linear perturbation theory by including overtones \cite{Giesler:2019uxc,Mourier:2020mwa}.  This is also related to the earlier work on the close-limit approximation which shows that it is possible to model the merger regime within linear perturbation theory \cite{Price:1994pm,Anninos:1995vf}.  On the other hand, there is also more recent evidence of the presence of quadratic combinations of the quasi-normal modes which would arise in second-order perturbation theory \cite{Cheung:2022rbm,Mitman:2022qdl,Khera:2023oyf} but these will be sub-dominant and for observational purposes we can consider linear perturbation theory to be a good approximation a short time after the merger.  It is not clear when (and if \cite{Thrane:2017lqn}) this perturbative regime can be distinguished and the choice of start time is an important issue.  See
\cite{Kamaretsos:2011um,Bhagwat:2017tkm} for studies of the start time
of the ringdown phase, and \cite{Okounkova:2020vwu,Mitman:2022qdl,Lagos:2022otp,Cheung:2022rbm}
for studies of possible non-linear effects.  The different regimes seen in the gravitational wave signal are expected to have counterparts in the strong field 
dynamical spacetime region near the binary system \cite{Jaramillo:2012rr,Jaramillo:2011re,Prasad:2020xgr}. 
See e.g. \cite{Mourier:2020mwa,Forteza:2021wfq,Pook-Kolb:2020jlr,Gupta:2018znn,Chen:2022dxt}
for studies of black hole horizon geometry in the post-merger phase
and whether a ringdown regime can be identified using the horizon
dynamics as well.

It has long been expected that only the most dominant ringdown mode
will be observable with the current generation of gravitational wave
detectors \cite{Berti:2016lat,Cabero:2019zyt}. Those expectations were
based on astrophysical assumptions about the total mass and mass ratio
distributions of binary black hole systems in the observable
universe, which in turn determine the amplitudes of various ringdown modes \cite{Borhanian:2019kxt}. However, evidence for an overtone of the dominant mode of
GW150914 was presented in \cite{Isi:2019aib,Giesler:2019uxc}. There it was shown that it is possible to model the
gravitational waveform as a superposition of ringdown modes
starting from the merger by using the overtones of the dominant mode.  This is a significant result, though there remain several interesting open questions regarding data analysis and theoretical issues. 
Some of the data analysis issues are discussed in
\cite{Cotesta:2022pci,Isi:2022mhy,Isi:2023nif,PhysRevLett.131.169001,PhysRevLett.131.169002}.  On the theoretical side, the stability of the overtones  under small perturbations raises several interesting open questions; see e.g. \cite{Nollert:1996rf,Nollert:1998ys,Jaramillo:2020tuu,Jaramillo:2021tmt,Destounis:2021lum}.
Evidence of a second fundamental mode, without using overtones, was
first presented for the event GW190521 in \cite{Capano:2021etf}, henceforth referred to as ``Capano et al.", and will be elaborated further in this paper.

With this analysis we address three fundamental questions. Firstly, if a signal explicitly does not contain any sub-dominant ringdown modes, how often does our detection pipeline falsely claim the existence of such modes? Secondly, if one or more sub-dominant modes are present in the data, how often does our pipeline correctly recover them? Thirdly, if our pipeline is used to constrain deviations from Kerr, how well do the resulting inferred parameters match those of the simulated signal? The key results for detection of a second mode are shown in Figs.~\ref{fig:pvalue_vs_zeta}~and~\ref{fig:cumbfs}. Figure~\ref{fig:pvalue_vs_zeta} shows the result of applying the ``agnostic analysis'' performed in Capano et al. to a set of simulated signals that do not have a $(3,3,0)$ mode as compared to a set that do. Figure~\ref{fig:cumbfs} (left) applies the ``Kerr analysis'' from Capano et al. to simulated signals without a $(3,3,0)$ mode, and shows that the false alarm probability is consistent with expectations from noise. Figure~\ref{fig:cumbfs} (right) quantifies the ability of the Kerr analysis to detect the $(3,3,0)$ mode when it is present, as a function of Bayes factor.

In section \ref{sec:ringdownPE} we give additional details of how the data is treated in the analysis of Capano et al. Section~\ref{sec:injectionsets} explains how we generate the simulated data sets.  
In sections \ref{sec:statsig_agnostic} and \ref{sec:statsig_Kerr} we investigate the statistical significance of detecting two modes versus one using a set of simulated signals. Section \ref{sec:statsig_agnostic} presents an agnostic analysis that looks at the consistency of the second mode with the first mode. Section \ref{sec:statsig_Kerr} presents an analysis more closely tied to the Kerr hypothesis, analysing the likelihood of two Kerr modes versus just one. In section \ref{sec:testGR} we use our simulated signals to compare the accuracy with which the no hair theorem can be tested using fundamental modes for an event similar to GW190521.

We conclude this introduction by briefly summarizing some basic properties of the event GW190521, which will be relevant in the rest of this paper.

\subsection{GW190521}

The gravitational wave event GW190521 was detected on May 21st 2019 at 03:02:29 UTC by the Advanced LIGO and Advanced Virgo detectors \cite{LIGOScientific:2020iuh}. The most conservative explanation of the signal is the binary merger of two black holes \cite{LIGOScientific:2020iuh,LIGOScientific:2020ufj}, although there are also various other interpretations of this event \cite{Bustillo:2020syj,Abedi:2021tti,Wang:2021gqm,Gamba:2021gap,DallAmico:2021umv,Shibata:2021sau}. 

While the progenitors of the event GW190521 are open to speculation, in most scenarios the final outcome is still likely to be a single black hole. The event GW190521 shows clear evidence of a dominant ringdown mode of a final black hole after the merger \cite{LIGOScientific:2020iuh}. In Capano et al.~\cite{Capano:2021etf} the ringdown signal was found to contain an additional sub-dominant ringdown mode. The dominant mode is consistent with being the $(\ell,m,n)=(2,2,0)$ ringdown mode of a Kerr black hole; the second mode is consistent with the sub-dominant fundamental $(\ell,m,n)=(3,3,0)$ mode. As detailed in this paper, under a Kerr hypothesis, the Bayes factor preferring the existence of the $(2,2,0)$ and $(3,3,0)$ modes over just the $(2,2,0)$ or the $(2,2,0)$ and $(2,2,1)$ modes is estimated to be $\eventbf{}$ (1$\sigma$ uncertainty).

If GW190521 is indeed a binary black hole merger, the inferred total mass of the system would make it one of the most massive binary black hole systems observed to date \cite{LIGOScientific:2021djp,Nitz:2021zwj}. Other interpretations have found even higher total masses \cite{Bustillo:2020syj}. A high total mass implies that very little of the inspiral phase occurs inside the sensitive band of the detectors and the recorded signal is dominated by the merger and ringdown.  Therefore an analysis that focuses solely on the ringdown phase is of interest and avoids some of the modelling issues in the progenitor inspiral phase.

\begin{figure}
    \includegraphics[width=\columnwidth]{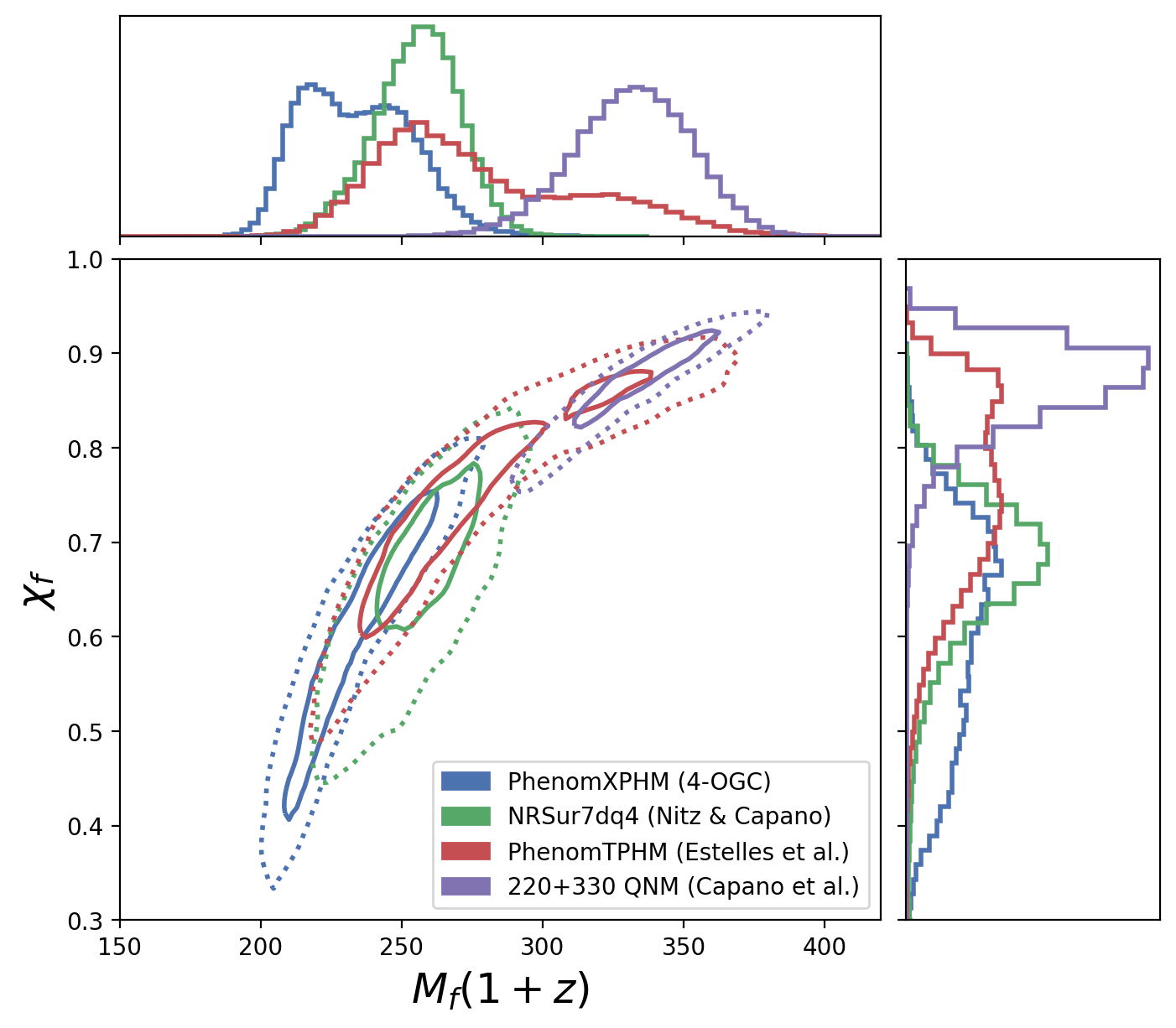}
    \caption{Comparison of the final mass and spin of GW190521, as estimated by NRSur7dq4~\cite{Nitz:2020mga}, IMRPhenomXPHM~\cite{Nitz:2021zwj}, IMRPhenomTPHM~\cite{Estelles:2021jnz}, and a Kerr ringdown with both the $(2,2,0)$ and $(3,3,0)$ modes~\cite{Capano:2021etf}. 
     {{The solid and dashed lines correspond to $50\%$ and $90\%$ credible contour, respectively.}}
     The IMRPhenomTPHM results have a second mode in the posterior that is consistent with the Kerr ringdown results.
   }
    \label{fig:imrcompare-final_mass_spin}
\end{figure}

\begin{figure}
    \includegraphics[width=\columnwidth]{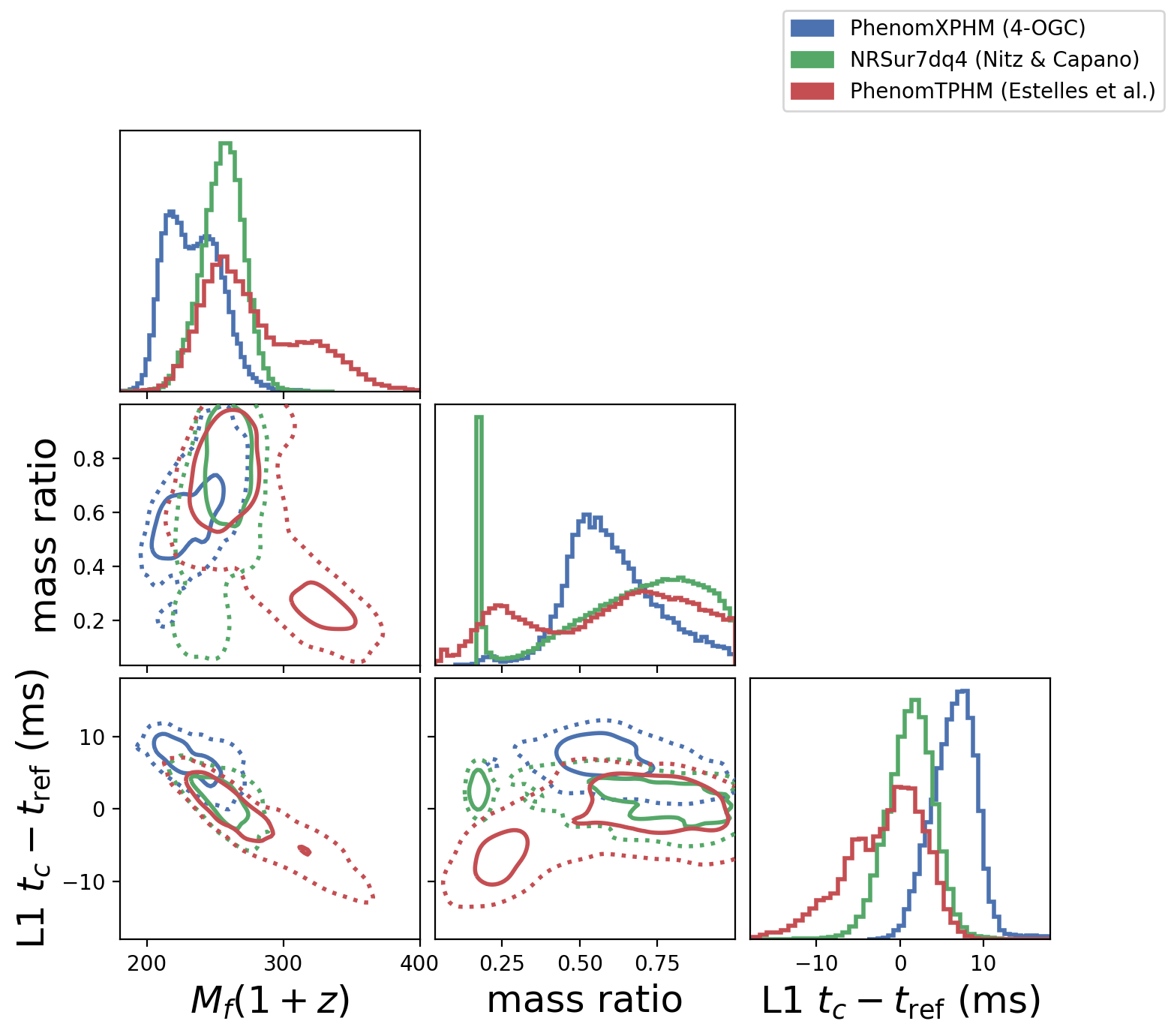}
    \caption{Comparison of the final mass, mass ratio, and coalescence time in the Livingston detector as estimated by the NRSur7dq4~\cite{Nitz:2020mga}, IMRPhenomXPHM~\cite{Nitz:2021zwj}, and IMRPhenomTPHM~\cite{Estelles:2021jnz} waveform models. 
    {{The solid and dashed lines correspond to $50\%$ and $90\%$ credible contour, respectively.}}
    The second mode in final mass found by IMRPhenomTPHM, which is consistent with the Kerr ringdown results (cf.~Fig.~\ref{fig:imrcompare-final_mass_spin}), corresponds to a second mode at more asymmetric masses. This mode also yields a coalescence time that is $\sim 6\,$ms earlier than the equal mass mode found by the other approximants. This earlier coalescence time estimate is $\sim 12\,$ms before the time at which the Bayes factor for the $(2,2,0)+(3,3,0)$ mode peaks in Capano et al.~\cite{Capano:2021etf}. It is also consistent with the time at which the evidence for the $(2,2,0)+(2,2,1)$ Kerr ringdown model peaks.
    }
    \label{fig:imrcompare-time_mass_ratio_final_mass}
\end{figure}

Inferences about the final black hole parameters using the ringdown signal alone are sensitive to the assumed start time of the ringdown \cite{Kamaretsos:2011um,Cotesta:2022pci}. Different starting times can lead to different results \cite{Cabero:2017avf,Kastha:2021chr}. A ringdown-only analysis must explicitly exclude some of the signal that is outside the ringdown phase. In this work we present additional details of the approach used in Capano et al. Parameter estimates for the event GW190521 based on the binary black hole interpretation are shown in Figs.~\ref{fig:imrcompare-final_mass_spin} and \ref{fig:imrcompare-time_mass_ratio_final_mass}. These estimates come from different authors using different methods \cite{Nitz:2020mga,Estelles:2021jnz,Capano:2021etf,Nitz:2021zwj}. The redshifted final total mass spans a wide range from around $200~ \mdot$ to nearly $400~\mdot$.

The peak gravitational wave strain is expected to occur close to the merger. 
The GPS time of the peak strain in the Hanford detector was initially estimated by the LIGO Scientific and Virgo Collaborations (LVC) using a numerical relativity (NR) surrogate waveform model, NRSur7dq4 \cite{Varma:2019csw}, to be $1242442967.4306^{+0.0067}_{-0.0106}$ (median $\pm 90\%$ credible interval) \cite{LIGOScientific:2020ufj}. See \cite{Estelles:2021jnz} for further discussion. 
As can be seen in Fig.~\ref{fig:imrcompare-time_mass_ratio_final_mass}, estimates of the merger coalescence time range over some $20\,$ms depending on the waveform model considered. This is a significant time range since, in geometric units, it corresponds to approximately $13M$ for an object with a mass $M=300~\mdot$. For this reason, in Capano et al. a range of starting times for the ringdown analysis was used, which spans the uncertainty of merger time estimation. See Section \ref{sec:injectionsets} for more details.

Subsequent to Capano et al. and this work, Siegel et al.~\cite{Siegel:2023lxl} found evidence for sub-dominant ringdown modes in GW190521. The identification of the modes differed from Capano et al. In the Siegel et al. analysis the $(2,1,0)$ and $(2,2,0)$ modes were most prominent, while the $(3,3,0)$ mode identified in Capano et al. was labelled as the $(3,2,0)$ mode. This yielded mass and spin estimates that were more compatible with the NRSur7dq4 results (a major motivation for their work). However, the amplitude of the $(2,1,0)$ mode in their model exceeds that of the $(2,2,0)$. This can happen for highly precessing signals~\cite{Zhu:2023fnf}. In Capano et al. the amplitude prior on all modes were constrained to be less than that of the $(2,2,0)$. We use the same priors here, and focus our attention on the $(2,2,0)$, $(3,3,0)$, and $(2,2,1)$ modes, relevant to the results in Capano et al. Consequently, our results here are not designed to discriminate between the interpretation of Capano et al. and the interpretation presented in Siegel et al.

\section{Basics of ringdown detection and parameter estimation}
%***************************************************
\label{sec:ringdownPE}

In this section we summarize some essential elements of the data
analysis procedure that we employ.  Since we analyze exclusively the ringdown which is only a part of the full signal, an important
challenge is to identify the portion of the full signal corresponding
to the ringdown.  Similarly, it is necessary to ensure that the
procedure for extracting this portion of the data properly takes
into account correlations with neighboring time samples that should be excluded from the analysis. 

Let $\vdata = \{\data_0, ..., \data_{N - 1}\}$ denote time-ordered samples
of the strain data from a gravitational wave detector.  The data is
sampled every $\Delta t$ seconds over a duration $T$, so that the
number of samples is $N = \lfloor T/\Delta t \rfloor + 1$.  A network of
$\numdet{}$ detectors sampled in this way will produce a set of
samples $\vdata_{\net} = \{\vdata_1, ..., \vdata_{\numdet}\}$.  The
strain data is assumed to be a combination of a possible signal
$\vsignal$ and noise $\vnoise$
\begin{equation}
  \vdata = \vsignal + \vnoise\,.
\end{equation}
Let $p(\vdata|\vec{\lambda},H)$ be the likelihood of the data $\vdata$ in the presence of a signal with given parameters $\vec{\lambda}$ under background hypotheses $H$, such as the signal model.
The probability of finding a realisation of noise $\vnoise$ under the hypotheses $H$ is $p(\vnoise | H)$.
Therefore the likelihood for data $\vdata$ can be written as
\begin{equation}
  p(\vdata|\vec{\lambda},H) = p(\vdata - \vsignal(\vec{\lambda})|H,n)\,,
\end{equation}
where the right-hand side is under the hypothesis $n$ that no signal is present.
In gravitational-wave astronomy, in the absence of a signal, it is common to assume over short times that the detectors output stochastic Gaussian noise
which is independent across detectors. With this assumption the probability density function describing the time-ordered noise samples of the detector network $\vnoise_{\net}$ is a product of $\numdet{}$ $N-$dimensional multivariate normal distributions,
\begin{equation}
\label{eqn:pnoise}
p(\vnoise_{\net}) = \frac{\exp\left[
    -\frac{1}{2}\sum_{d=1}^{\numdet{}} \vnoise_{d}^{\top}\vcovmat_{d}^{-1} \vnoise_{d}\right]}    {\sqrt{(2\pi)^{N\numdet{}} \prod_{d=1}^{\numdet{}}\det \vcovmat_{d}}}.
\end{equation}
Here, $\vcovmat_{d}$ is the covariance matrix of the noise in detector
$d$, and we drop the hypotheses $H$ in our notation. See \cite{Finn:1992wt} for further details.

If the detector's noise is
wide-sense stationary and ergodic, which is typically the case for the LIGO and Virgo detectors, the noise likelihood takes a simple form
\begin{equation}
    p(\vec{n}_\mathrm{net}) \propto \exp\left[-\frac{1}{2}\sum_{d=1}^{K} \ip{\vnoise_d}{\vnoise_d}\right].
\label{eqn:likelihood}
\end{equation}
Here, the inner product $\ip{\cdot}{\cdot}$ is defined as
\begin{equation}
\ip{\vec{u}_d}{\vec{v}_d} \equiv 4 \Re \left\{\frac{1}{T} \sum_{p=1}^{N/2-1} \frac{\tilde{u}_d^{*}[p]\tilde{v}_d[p]}{\psd^{(d)}[p]} \right\},
\label{eqn:innerprod}
\end{equation}
where $\tilde{u}$ is the discrete Fourier transform of the time series $\vec{u}$, an asterisk denotes the complex conjugate, and $\psd$ is the power spectral density of the detector's noise.
To obtain the posterior probability density function $p(\vec{\lambda}|\vdata, H)$ for the parameters $\vec{\lambda}$, we use Bayes' theorem
\begin{equation}
    p(\vec{\lambda}|\vdata, H) = \frac{1}{Z} p(\vdata|\vec{\lambda}, H) p(\vec\lambda|H),
\end{equation}
where $p(\vdata|\vec{\lambda}, H)$ is the likelihood function, $p(\vec\lambda|H)$ is the prior, and $Z$ is a normalization constant known as the evidence, depending only on the data. 
Taking the ratio of evidences $Z_A/Z_B$ for two different models $H_A$ and $H_B$ yields the ``Bayes factor''. 
In this work, the signal models will be GW ringdown waveforms with only fundamental modes and overtones.
If our prior belief for the validity of the two models is the same, the Bayes factor gives the odds that model A is favoured over model B. 
Ref.~\cite{Kass:1995loi} suggested Bayes factors greater than 3.2, 10 and 100 are considered substantial, strong, and decisive, respectively.

The ringdown waveform model takes the following form
\begin{equation}
\label{linear_spectrum}
h_+ + i h_\times = {\displaystyle \frac{M_f}{D_L}} \sum_{\lmn} {}_{_{-2}}S_{\lmn} (\iota, \varphi,{\chi}_f) 
A_{\lmn} e^{i(\Omega_{\lmn}t +  \phi_{\lmn})} \, ,
\end{equation}
where $h_{+/\times}$ are the plus/cross polarizations of the wave, $M_f$ is the total mass of the remnant black hole in the detector frame and $D_L$ is the source luminosity distance. The waveform is decomposed with respect to the spin-2 weighted spheroidal basis ${}_{_{-2}}S_{\lmn}$, which is a function of the remnant black hole's spin $\chi_f$, the inclination angle $\iota$ and azimuthal angle $\varphi$ relative to the observer.
The amplitude and phase of the quasi-normal modes are denoted by $A_{\lmn}$ and $\phi_{\lmn}$.
The complex frequency is $\Omega_{\lmn} = 2\pi f_{\lmn} + i /\tau_{\lmn}$, where the characteristic frequency $f_{\lmn}$ and decay time $\tau_{\lmn}$ are solely determined by the mass and spin of the remnant black hole, as predicted by the no-hair theorem in GR.
We also consider an agnostic ringdown waveform model in this work, for which we absorb the $M_f/D_L$ term into the amplitude and replace the spheroidal harmonics with arbitrary complex numbers $X_{\ell\pm mn}=e^{i\psi_{\ell \pm m n}}$.

In a standard full-signal analysis, to obtain the likelihood for the signal hypothesis,
the noise $\vnoise_d$ in Eq.~\ref{eqn:likelihood} is replaced by the residuals $\vdata_d - \vsignal_d$. 
This requires that $\vsignal$ is an accurate model of the signal across the entire observation time $T$, which is not valid for a ringdown-only analysis.
Quasi-normal modes only model the gravitational wave from a binary black hole after the merger, when the two component black holes have formed a single, perturbed black hole. 
Performing Bayesian inference using quasi-normal modes as the signal model therefore requires ignoring times from the data when the ringdown prescription is not valid.

We perform the ``gating and in-painting'' technique \cite{Zackay:2019kkv} to remove the influence of pre-ringdown data.
Define $\vnoise' = \vnoise_g + \vec{x}$, where $\vnoise_g$ is the noise with the pre-merger data zeroed out.
We solve for $\vec{x}$ such that $\vcovmat^{-1} \vnoise' = 0$ in the gated region. Doing so, we can use $\vnoise'$ in the frequency-domain likelihood equation Eq.~\ref{eqn:likelihood} to obtain the same result as if we excised the gated time and directly computed the likelihood in Eq.~\ref{eqn:pnoise}.

We use the gated-Gaussian likelihood described above in the open source \texttt{PyCBC Inference} library~\cite{pycbcgithub,Biwer:2018osg}. We evaluate the noise residuals with $\vnoise_g = \vdata_g - \vsignal_g$ (i.e., the residual with the gated region zeroed out) and solve for $\vec{x}$ under the condition
\begin{equation}
    \overline{\vcovmat^{-1}}\,\overline{\vec{x}} = -\overline{\vcovmat^{-1}\vnoise_g},
\label{eqn:gatecondition}
\end{equation}
where the overbar indicates the gated region. 
We can then use $\vec{x} + \vdata_g - \vsignal_g$ in the standard likelihood, Eq.~\ref{eqn:likelihood}. 

For all analyses we use a gate of two seconds, ending at the start time of the ringdown. 
We use the ``v2'' $4\,$kHz data for GW190521~\cite{gw190521data}, which is made publicly available by the Gravitational Wave Open Science Center~\cite{LIGOScientific:2019lzm}. For all injections we fix the sky location to the values given by the maximum likelihood result from the NRSur7dq4 analysis in Nitz \& Capano~\cite{Nitz:2020mga}. For sampling the parameter space we use the \texttt{dynesty} nested sampler~\cite{2020MNRAS.493.3132S}. 

In this work we consider a variety of signal models with different combinations of angular and overtone modes characterized by Eq.~\ref{linear_spectrum}.
The fundamental mode is $(\ell,m,n)=(2,2,0)$, and we further consider models with an additional $(2,2,1)$ overtone or $(3,3,0)$ mode, whose complex frequencies are either predicted by the Kerr hypothesis or treated agnostically as parameters to be determined.
We list the priors $p(\vec\lambda|H)$ for all parameters used in this work in Table~\ref{table:prior}.
In particular, the (2,2,1) amplitude is chosen to be [0,5] times that of the (2,2,0) mode's. This choice is motivated by the numerical relativity fits from Giesler et al. \cite{Giesler:2019uxc}, which found that the (2,2,1) mode could be $\sim4$ times louder than the (2,2,0).
For the $(3,3,0)$ amplitude we chose a prior that is $[0, 0.5]$ times that of the (2,2,0) mode, which is informed by the numerical simulation results of binary black hole mergers in Ref.~\cite{Borhanian:2019kxt}.  See also \cite{Kamaretsos:2011um,Li:2021wgz} which discuss amplitudes of the $(3,\pm 3,0)$ modes.

When sampling the posterior for the Kerr and agnostic analysis, we numerically marginalize the polarization angle using a discrete grid of 1000 points. The original motivation was to speed up sampler convergence for the large number of injections analyzed here. However, we found that doing so also led to more robust estimates of the Bayesian evidence, as the sampler was better able to converge on the posterior. Consequently we also reanalyzed GW190521 using the numerical marginalization of the polarization. The effect on the estimation of the Bayes factor is discussed in more detail in Appendix~\ref{app:polmarg}. 

\begin{table*}[t]
  \begin{center}
\begin{tabular}{|P{45pt}|P{50pt}|p{220pt}|p{120pt}|}
\hline
{\bf Model} & {\bf Parameter} & \hfil {\bf Parameter description} & \hfil {\bf Uniform prior range} \\
\hline
 & $f_{A/B/C}$ & frequencies of regions A/B/C &  $[50,80]$/$[80,256]$/$[15,50]$ Hz \\
& $\tau_{A/B/C}$ & decay times of regions A/B/C &  $[0.001,0.1]$ s \\
& $\log_{10} A_B$ & base-10 logarithm of the amplitude of region B &  [-24,19] \\
Agnostic &$A_{A/C} / A_B$ & ratio of amplitudes between region A/C and region B &  [0,0.9]\\
&$\phi_{A/B/C}$ & phases of regions A/B/C &  $[0,2\pi]$\\
&$\psi^{+/-}_{A/B/C}$ & phase of the +m and -m modes of the arbitrary complex number in region A/B/C &  $[0,2\pi]$\\
&$d\beta$ & angular difference in amplitudes of +m and -m modes  &  $[-\pi/4,\pi/4]$\\
\hline
&$M_f$ & final black hole mass in the detector frame & [100,500] $M_\odot$\\
&$\chi_f$ & final black hole spin &  [-0.99,0.99] \\
Kerr &$\log_{10} A_{220}$ & base-10 logarithm of the amplitude of (2,2,0) & [-24,-19] \\
&$A_{330/220}$ & ratio of amplitudes between $(3,3,0)$ and $(2,2,0)$  &  [0,0.5] \\
&$A_{221/220}$ & ratio of amplitude between $(2,2,1)$ and $(2,2,0)$  &  [0,5] \\
&$\phi_{220/330/221}$ & phase of (2,2,0)/(2,2,1)/(3,3,0) &  $[0,2\pi]$ \\
\hline
&$\delta f_{221}$ &  fractional deviation from GR of the (2,2,1) frequency &  [-0.16,0.3] with the constraint $f_{221}(1+\delta f_{221})>55$ Hz\\
No hair test &$\delta \tau_{221}$ & fractional deviation from GR of the (2,2,1) decay time &  [-0.8,0.8]\\
&$\delta f_{330}$ & fractional deviation from GR of the $(3,3,0)$ frequency  &  [-0.3,0.3] with the constraint $f_{330}(1+\delta f_{330})>75$ Hz\\
&$\delta \tau_{330}$ &fractional deviation from GR of the $(3,3,0)$ decay time  & [-0.9,3]\\
\hline
All models&$\cos\iota$ & cosine of inclination angle &  [-1,1] \\
&$\psi$ & polarization angle &  $[0,2\pi]$ \\
\hline
\end{tabular}
\caption{Prior distributions of sampling parameters for the models used in this work:
The agnostic model with the spheroidal harmonics replaced by an arbitrary complex number as discussed in Sec.~\ref{sec:statsig_agnostic}, the Kerr model described by Eq.~\ref{linear_spectrum} and discussed in Sec.~\ref{sec:statsig_Kerr}, and the testing-GR model discussed in Sec.~\ref{sec:testGR}.}
  \label{table:prior}
  \end{center}
\end{table*}

\section{Selection of simulated signals}
\label{sec:injectionsets}
%************************************************

In this paper we seek to validate the evidence for the observation of the $(3,3,0)$ mode in GW190521. To do so, we create two sets of simulated signals (``injections''): one set with no $(3,3,0)$ mode in the ringdown (the \emph{\controlinj{}} set), and another set containing a $(3,3,0)$ mode in the ringdown (the \emph{\signalinj{}} set). The \controlinj{} set is used to measure the rate of false alarms -- i.e., to answer the question, how often do we get large evidence for the $(3,3,0)$ mode when the signal contains no $(3,3,0)$ mode? -- while the \signalinj{} set is used to validate that our pipeline can in fact detect a $(3,3,0)$ mode when it exists in the signal.

For the \controlinj{} injections we randomly select 500 points from the NRSur7dq4 posterior published in Nitz \& Capano~\cite{Nitz:2020mga}. This posterior was similar to the posterior published in the initial LIGO/Virgo publication on GW190521~\cite{LIGOScientific:2020ufj}. With the exception of a secondary peak in the posterior around $m_1/m_2 \sim 6$, this NRSur7dq4 posterior favored approximately equal masses for the binary.\footnote{In Nitz \& Capano a prior uniform in $m_1/m_2 \in [1,6]$ was used in the NRSur7dq4 analysis. If a prior uniform in $m_2/m_1$ is used (which is approximately the same as a prior uniform in component masses, as done in the LIGO/Virgo analysis), the second mode in the posterior at $m_1/m_2\sim6$ is down-weighted, giving further support to the equal-mass portion of the posterior. Here, we draw from the original posterior published in Nitz \& Capano, which used a prior uniform in $m_1/m_2$.} It also yielded a merger time for GW190521 only $\sim6\,$ms before the claimed observation time of the $(3,3,0)$ mode in Capano et al.\ and a relatively low final mass estimate; see Figs.~\ref{fig:imrcompare-final_mass_spin} and \ref{fig:imrcompare-time_mass_ratio_final_mass}. These results contrast with the claimed observation in Capano et al.: a large $(3,3,0)$ amplitude is not expected for equal-mass binaries~\cite{Borhanian:2019kxt}, and a ringdown model consisting of only fundamental modes is not expected to be a good model for the signal until $\sim10\,M$ after merger \cite{Kamaretsos:2011um}, which for GW190521 would be $\sim12-16\,$ms, not $\sim6\,$ms. As such, these injections are ideal to test the false alarm rate of our analyses.

To ensure that no $(3,3,0)$ mode exists in the \controlinj{} injections, we constrain all 500 injections to have mass ratios $m_2/m_1>0.5$ and we turn off all but the $\ell=2$ modes when generating the simulated waveforms. The waveforms are generated using the NRSur7dq4 approximant~\cite{Varma:2019csw}. We use 500 injections to get a sufficient number of samples at the Bayes factor of GW190521 (\eventbf{}); see Sec.~\ref{sec:statsig_Kerr} for more details.

To produce the \signalinj{} injections we draw random samples from the posterior published in Estelles et al.~\cite{Estelles:2021jnz}. This analysis used the IMRPhenomTPHM approximant to analyze GW190521. As with the results presented in Nitz \& Capano \cite{Nitz:2020mga}, Estelles et al.\ found a bimodal posterior in the component masses for GW190521: one mode favoring nearly equal masses, and one mode favoring mass ratios of $\sim4:1$. Intriguingly, as shown in Figs.~\ref{fig:imrcompare-final_mass_spin} and \ref{fig:imrcompare-time_mass_ratio_final_mass}, the second mode yielded a mass and spin estimate for the final black hole that is consistent with the estimate from the ringdown analysis in Capano et al. The estimated merger time for this second mode was also $\sim5-10\,$ms earlier than the NRSur7dq4 estimate, which is consistent with the peak in the $(2,2,1)$ Bayes factor found in Capano et al.\ and $\sim10\,M$ before the peak in the $(3,3,0)$ Bayes factor. The IMRPhenomTPHM waveforms are therefore ideal for our \signalinj{} injection set, particularly those from the more asymmetric mass ratio part of the posterior. 

Note that we do not assume this specific waveform model to be an exact representation of general relativity. For example, IMRPhenomTPHM is calibrated to aligned-spin simulations, relying on a ``twisting-up'' procedure that may be less reliable late in the inspiral~\cite{Estelles:2021jnz}. We validate that when higher order mode content is injected from waveform models with parameters which give the best available fit to GW190521, the presence or lack of higher order modes in these injections can be determined using our analysis framework.  We assume that the waveforms are sufficiently accurate to be useful for this purpose.

To try to ensure that the \signalinj{} injections have an observable $(3,3,0)$ mode after the merger, we draw 100 injections from the IMRPhenomTPHM posterior published in Estelles et al.\ and keep only those that have an estimated $(\ell,m,n) = (3,3,0)$ amplitude $>0.2$ after merger. We also require that the signal-to-noise ratio (SNR) of the $(3,3,0)$ mode be at least $4$ (the SNR estimated for the $(3,3,0)$ mode in GW190521 in  Capano et al.) at some point after merger. To estimate the $(3,3,0)$ SNR we filter each injection in noise with a template consisting only of the $(\ell, m)=(3,3)$ mode, and we gate both the template and signal to remove pre-merger times. Note that here, $(\ell,m)$ refer to \emph{spherical} harmonics, which is the basis used for inspiral-merger-ringdown (IMR) models, not the \emph{spheroidal} harmonics used for QNMs. Many of the posterior samples have large precession. Precession mixes the $m$ modes with the same $\ell$ in the observer frame. Consequently, an $(\ell,m) = (3,3)$ mode for a IMRPhenomTPHM waveform may consist of a combination of $(\ell, m, n)$ QNM modes, and not necessarily just the $(3,3,0)$ mode. As such, the estimated SNR may be considered an upper bound on the underlying $(3,3,0)$ QNM.

Applying the SNR cut to the initial 100 draws yields \numhiqinj{} \signalinj{} injections. We do not try to generate more \signalinj{} injections as they are only used to check that the analysis can recover signals with a $(3,3,0)$ mode and not to estimate small false alarm rates, as we do with the \controlinj{} injections. We use IMRPhenomTPHM to generate the waveform for the \signalinj{} set. Due to the differences between spherical and spheroidal modes, and to try to simulate a realistic signal, we use all available modes in IMRPhenomTPHM when generating the \signalinj{} set.

Both sets of injections are added to detector data at random times surrounding the estimated merger time of GW190521. Specifically, an offset time $t_{\rm offset}$ is drawn uniformly in $\pm[4, 20]\,$s and added to the coalescence time $t_c$ that is drawn from the relevant posterior for each injection. The gap of $\pm4\,$s around GW190521 is to prevent contamination of the data from GW190521. As described below, we perform ringdown analyses on a grid of times surrounding each injection. The widest grid -- used in the validation of the Kerr Bayes factor (see Sec.~\ref{sec:statsig_Kerr}) -- is $[-9,24]\,$ms. We therefore draw the $t_{\rm offset}$ such that they are at least $33\,$ms apart, to ensure that no two analyses analyze exactly the same detector data. 

As with the analysis of GW190521 in Capano et al., we use a reference time $t^{\rm inj}_{\rm ref}$ for each injection, around which we construct the grid of times used in the ringdown analyses. For each injection we set the reference time to be $t^{\rm inj}_{\rm ref} = t_{\rm ref} + t_{\rm offset}$, where $t_{\rm ref}=1242442967.445$ GPS seconds is the estimated geocentric merger time of GW190521, as determined by the maximum likelihood parameters taken from the NRSur7dq4 analysis in Nitz \& Capano~\cite{Nitz:2020mga}. This is the same $t_{\rm ref}$ used in Capano et al. Note that $t^{\rm inj}_{\rm ref}$ is not the injection's coalescence time $t^{\rm inj}_c$; instead, $t^{\rm inj}_c-t^{\rm inj}_{\rm ref}$ follow the same distribution as $t_c-t_{\rm ref}$ (see Fig.~\ref{fig:imrcompare-time_mass_ratio_final_mass}). We do this because our goal is to replicate the GW190521 analysis performed in Capano et al., including all uncertainties, so that our derived statistics are robust. Basing grid times on $t^{\rm inj}_c$ instead of $t_{\rm ref}^{\rm inj}$ would mean assuming information we did not know for the real signal. In the case of IMRPhenomTPHM, this means that some of our \signalinj{} injections merge as much as $20\,$ms before the reference time, well before the grid times used for the analysis.

\section{Statistical significance of the agnostic analysis}
%***************************************************
\label{sec:statsig_agnostic}

\begin{figure}
    \centering
    \includegraphics[scale=0.5]{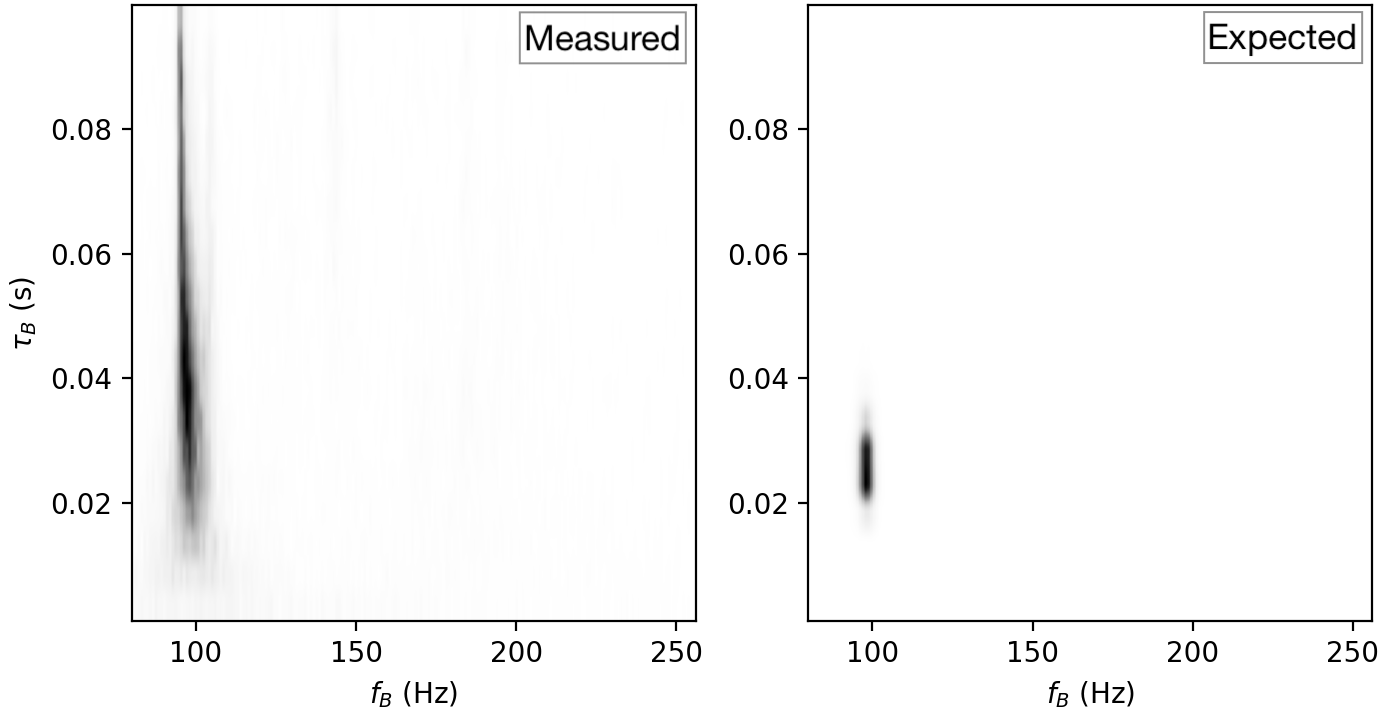}
    \caption{\textit{Left:} Marginal posterior of the frequency and damping time in frequency range B from the agnostic analysis of GW190521 at $t_{\rm ref} + 6\,$ms (same as the heat map in Fig.~1 of Capano et al). \textit{Right:} the expected distribution of the $(3,3,0)$ mode assuming the mode observed in frequency range A is the dominant, $(2,2,0)$ mode (same as the blue contour in Fig.~1 of Capano et al.) of a Kerr black hole. Darker regions indicate higher probability. The expected distribution is more concentrated due to the larger SNR of the dominant mode. We quantify the agreement between the measured and expected by multiplying the two distributions together and integrating ($\zeta$). The figure and $\zeta$ values are produced using 100 bins each in frequency and damping time.}
    \label{fig:agnosticGW190521}
\end{figure}

Two ringdown analyses of GW190521 were presented in Capano et al.~\cite{Capano:2021etf}: an ``agnostic'' analysis and a ``Kerr'' analysis. In the former, the data were analyzed using three QNMs with no assumption made about the relationship between the frequency and damping times of each mode. To prevent all three modes from locking on to the single dominant mode, each mode was assigned a separate frequency range: 50 -- 80Hz (range ``A''), 80 -- 256Hz (range ``B''), and 15 -- 50Hz (range ``C''). Range A covered the dominant mode, which was clearly visible in the data. This implicitly assumes that there is only one single mode in each range. The analysis was repeated in intervals of 6ms, between $t_{\rm ref} + [0, 24]\,$ms.

A signal with a well-defined posterior was found in Range A, having frequency $\sim 63$ Hz and damping time $\sim 26$ ms (see Fig.~1 of Capano et al.). No signal was found in Range C. A second putative mode was found in Range B. This signal was most pronounced at $t_{\rm ref} + 6\,$ ms, at which point it has a frequency of $\sim 98$ Hz and damping time $\sim 40$ ms. As shown in Fig.\ref{fig:agnosticGW190521}, these frequencies and damping times were where one would expect the $(3,3,0)$ would be assuming the remnant of GW190521 is a Kerr black hole, with the signal in Range A being the $(2,2,0)$ mode.

Initially, the agnostic analysis was presented as qualitative evidence for the presence of the $(3,3,0)$ mode. Here, we repeat the analysis on our two sets of injections and use them to develop a statistic to quantify the statistical significance of the agnostic result. Our aim is to find a statistic that can separate the \controlinj{} injections from the \signalinj{} injections.

If an observable $(3,3,0)$ mode is truly present in the signal, and it is the only observable mode in Range B, then the measured posterior distribution should peak at the same values as the expected distribution. We expect the measured distribution to be more diffuse than the expected distribution. This is because the expected distribution is derived from the observed dominant mode, which is more accurately measured due to its larger SNR. With these considerations in mind, we quantify the agreement between the measured distribution $p_{\rm meas}(f_B, \tau_B)$ and the expected distribution $p_{330}(f_B, \tau_B)$ using:
\begin{equation}
    \zeta \equiv \int p_{\rm meas}(f_B, \tau_B) p_{330}(f_B, \tau_B) \mathrm{d}f_B \mathrm{d}\tau_B.
    \label{eqn:zetadef}
\end{equation}
This is effectively an inner product between the measured and expected distributions. To evaluate this we construct 2D histograms in Range B using 200 bins in frequency and 20 bins in damping time. This is done at each time step; we then maximize $\zeta$ over all the time steps.

\begin{figure}
    \centering
    \includegraphics[width=\columnwidth]{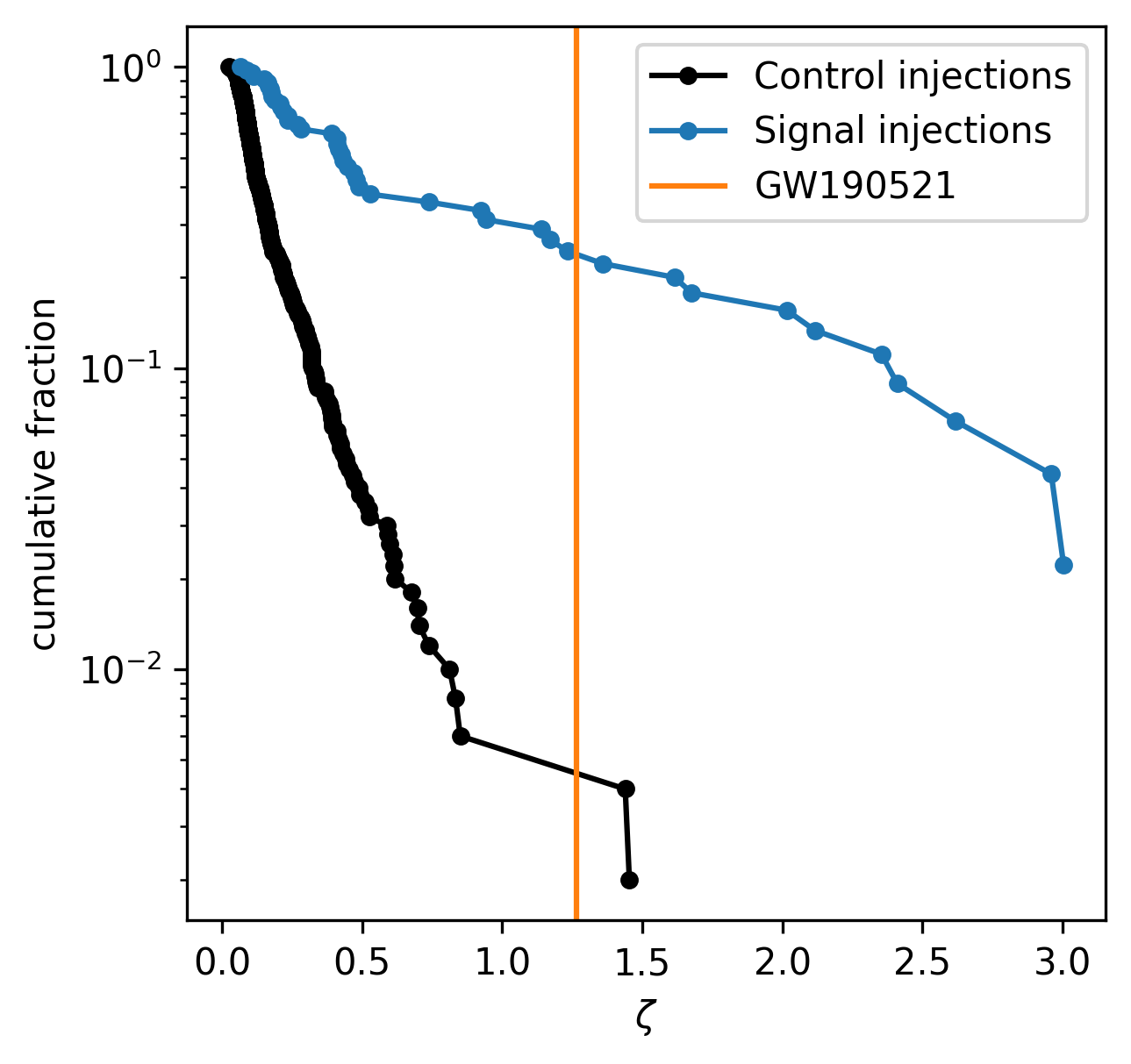}
    \caption{Cumulative distribution of $\zeta$ values for the \controlinj{} injections (black dots/line) and the \signalinj{} injections (blue dots/line). The orange vertical line shows $\zeta$ for GW190521. We use the cumulative distribution of \controlinj{} injections to estimate a p-value for the $\zeta$ of GW190521. Since two of the 500 \controlinj{} injections have a $\zeta$ larger than GW190521, we estimate the p-value of the agnostic result to be 0.004}
    \label{fig:pvalue_vs_zeta}
\end{figure}

We calculate $\zeta$ for the \controlinj{} injections. Since these injections have no $(3,3,0)$ mode by construction, the resulting $\zeta$ values represent the distribution of false positives. The cumulative distribution of the maximized $\zeta$ is shown by the black line in Fig.~\ref{fig:pvalue_vs_zeta}. We also calculate $\zeta$ for the \signalinj{} injections that have post-merger SNR $>$ 4. The cumulative distribution of the maximized $\zeta$ is also shown in Fig.~\ref{fig:pvalue_vs_zeta} as a blue line. As evident in the figure, we find good separation between the signal and control injections. 

Calculating $\zeta$ for GW190521, we find that it is at a maximum at $t_{\rm ref} + 6\,$ms, with a value of 1.27. This is consistent with our initial qualitative assessment that the observed mode is most consistent with the expected $(3,3,0)$ mode at $6\,$ms. As shown in Fig.~\ref{fig:pvalue_vs_zeta}, only two of the 500 \controlinj{} injections have a $\zeta$ larger than GW190521. We therefore conclude that the probability of obtaining a $\zeta$ greater than or equal that of GW190521 by chance from noise is $0.004$. Note that this statistic does not indicate whether there is the presence of a second mode in range B, but whether the mode in range B is consistent with the mode in range A, given Kerr assumptions.

\section{Statistical significance of the Kerr analysis}
%***************************************************
\label{sec:statsig_Kerr}

In the Kerr analysis we assume the final black hole is described by the Kerr metric. In this case, the frequencies and damping times of all post-merger QNMs are given uniquely by the mass and spin of the black hole. Each additional mode therefore adds two additional degrees of freedom: one for the amplitude and one for the phase of the mode. The relative amplitudes and phases of the modes can in principle be determined by the pre-merger component masses, their spins, and their relative orientation at merger.

Knowing what amplitude and phase to use for each mode requires detailed knowledge of the pre-merger conditions, which are not easily discernible for events like GW190521 in which the pre-merger signal is short and difficult to observe. Furthermore, models mapping pre-merger properties to post-merger quasi-normal modes (QNMs) are limited for highly precessing systems, particularly those with large ($\gtrsim 2$) mass ratios. For these reasons, even when assuming a Kerr model for the post-merger signal, we use uniform priors on the phases and relative amplitudes of the sub-dominant modes with respect to the dominant mode.

Using such broad priors on amplitude and phase makes the analysis susceptible to overfitting. In principle, all modes are present in the signal. However, the vast majority of these modes are negligible compared to the dominant mode. For the types of signals detectable by the current generation of detectors, we expect only a few fundamental modes to have amplitudes that are at most $O(10\%)$ of the dominant mode's \cite{Berti:2016lat,Cabero:2019zyt,Zhu:2023fnf}. A signal model that contains more than a few modes that have broad amplitude priors is effectively unphysical (assuming the signal is sufficiently close to GR), and would yield lower Bayes factors due to Occam's razor. To give meaningful results, the signal model should only include the \emph{observable} modes, not the possible ones.

As with the agnostic analysis, the Kerr analysis also needs to determine when the observable modes are present. Before the merger the QNM model is not valid -- there is not a single perturbed black hole at this point. During the merger there may be non-linear components to the signal and/or significant contributions from overtones. Too late after the merger, and the signal will have damped away too much to make anything but the dominant mode observable.

We address both challenges through the use of Bayes factors. Given a signal model with observable modes $\vec{X}=\{(2,2,0), ...\}$ at a given time $t-t_{\rm ref}$, we calculate the evidence that the data contain those modes at that time,
\begin{equation}
Z_{\vec{X}}(t) = \int p(\vdata | \{\vec{\lambda}_{\vec{X}};t\}, \signal) p(\{\vec{\lambda}_{\vec{X}};t\} | \signal)\mathrm{d}\vec{\lambda}_{\vec{X}}.
\end{equation}
Taking the ratio of this evidence to the evidence for the $(2,2,0)$-only model ($Z_{220}$) at the same time gives the relative odds (or Bayes factor) that the data favor that model as compared to the $(2,2,0)$-only model.

As discussed above, we do not normalize the likelihood function in our analysis. This means that evidence values at different times cannot be directly compared to each other. However, the likelihood function's normalization factor cancels in the Bayes factor since the normalization only depends on the noise properties and not the signal model. It is therefore possible to compare Bayes factors at different times. 

Taking the point that $Z_{\vec{X}}/Z_{220}$ is at a maximum yields the time that the model with modes $\vec{X}$ is the best fit to the data relative to the $(2,2,0)$ model. However, the $(2,2,0)$-only model is known not to be a good model for the signal at merger~\cite{Giesler:2019uxc}. As a result, if we find $Z_{\vec{X}}/Z_{220}$ to be large at some time, it is not clear if this is because modes $\vec{X}$ are a good model for the signal, or if the $(2,2,0)$-only model is just a very bad model at that time. Put another way, $Z_{\vec{X}}/Z_{220}$ only tells us whether the $\vec{X}$ modes are a better fit for the data than just the $(2,2,0)$, not whether the $\vec{X}$-modes are truly observable. This problem becomes particularly acute as we get close to merger.

To account for this, we make use of the observation in Ref.~\cite{Giesler:2019uxc} that including overtones of the dominant mode better fit the signal close to (or even at) merger than the $(2,2,0)$-mode only. We modify the Bayes factor to be
\begin{equation}
    \label{eqn:bayes_factor}
    \bayesfac(X, t) \equiv \frac{Z_{X}(t)}{\max\{Z_{220}, Z_{220+221}\}}
\end{equation}
for all models $X\neq (2,2,0)+(2,2,1)$ (for the $(2,2,0)+(2,2,1)$ model we simply use $\bayesfac=Z_{220+221}/Z_{220}$). This allows us to both identify the most likely observable modes and the time at which they are most observable.

When applying this method to GW190521 we find \bayesfac $(220+330)$ to peak at \eventbftime ~with a value of \eventbf. This means that the $(2,2,0)+(3,3,0)$ model is \approxebf{} times more likely to be true than the $(2,2,0)$-only model, qualifying it as ``strong'' evidence for the $(3,3,0)$ model relative to the $(2,2,0)$-only model. In other words, if the signal did not have an observable $(3,3,0)$ mode, then we should expect to get a \bayesfac{} as large as this from noise only 1 in \approxebf{} times. 

To test the validity of this observation, we repeat the Kerr analysis on our \controlinj{} injections. As with GW190521, we repeat the analysis on a grid of times spanning $t_{\rm ref}^{\rm inj} + [-9, 24]\,$ms, although to reduce computational cost for the large number of analyses involved, we sample in intervals of $3\,$ms instead of the $1\,$ms interval used in Capano et al. Since our \controlinj{} injections contain no $(3,3,0)$ mode by construction, any large $\bayesfac$ observed with them is a false alarm. If our analysis assumptions are correct -- that the real data is Gaussian and that we are after the merger -- then on average we expect to get a $\bayesfac \geq \approxebf$ from $\expectedFA$ of the 500 injections.

\begin{figure*}[ht]
\centering
\includegraphics[width=\columnwidth]{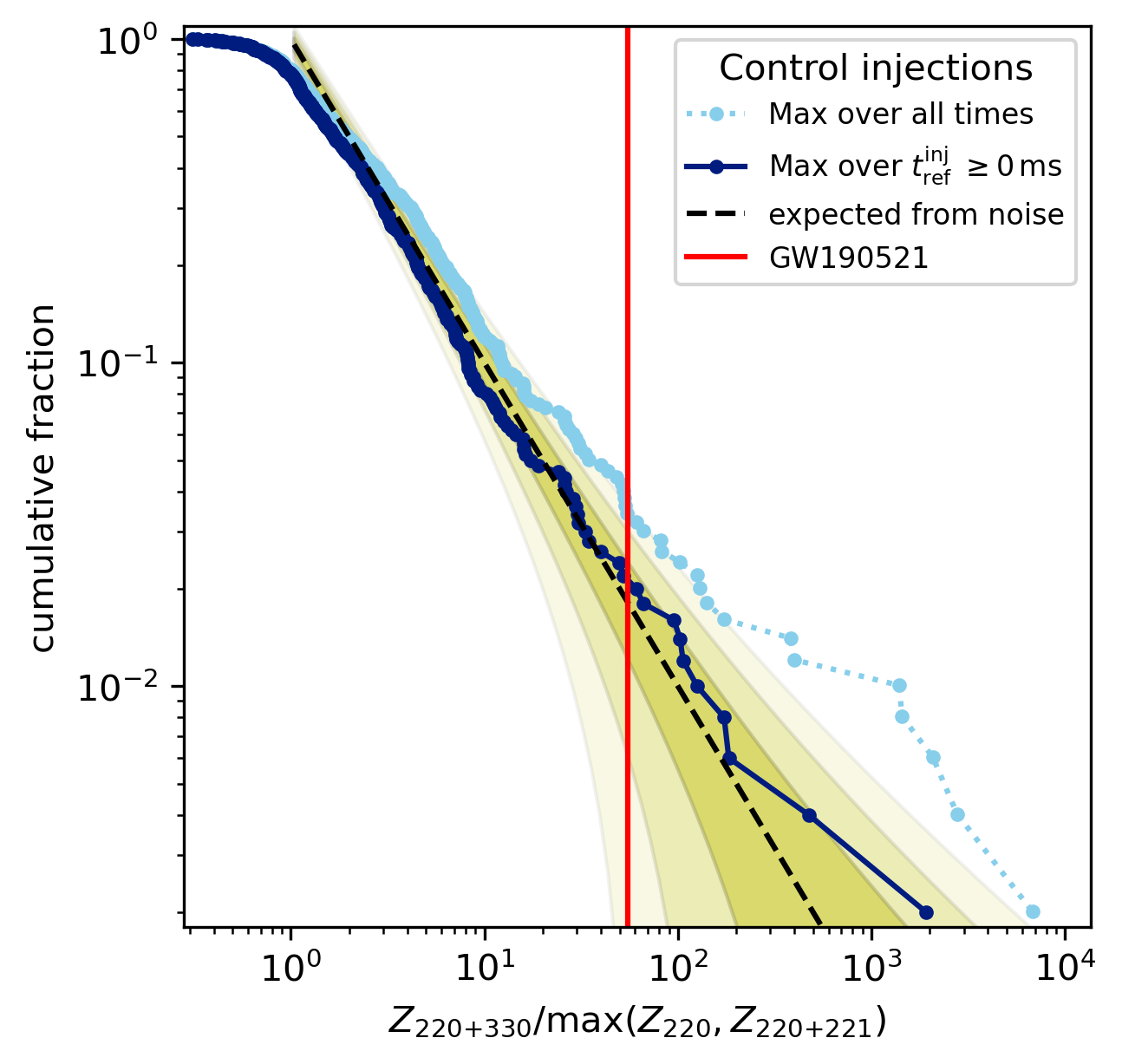} \hfill
\includegraphics[width=\columnwidth]{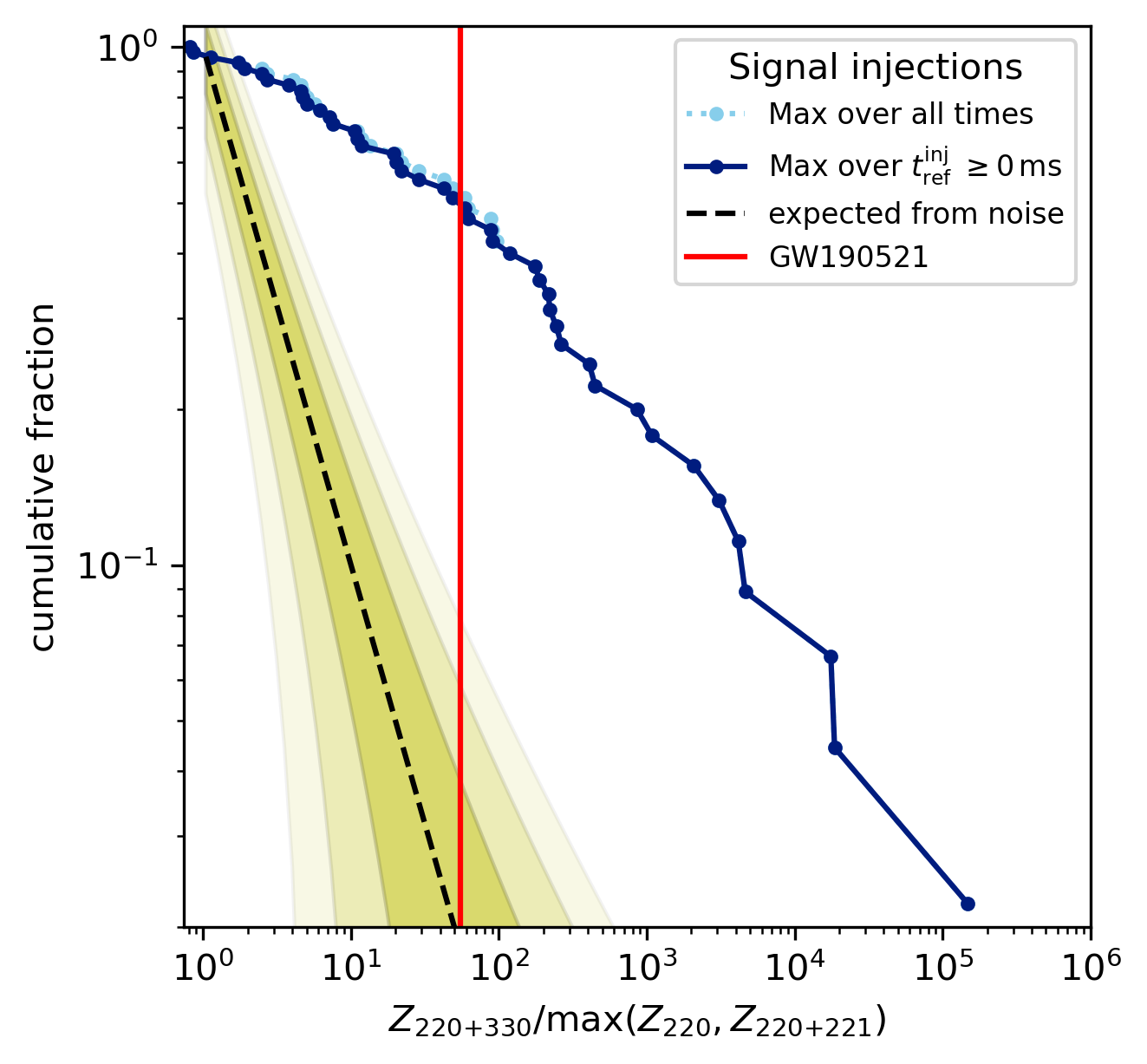}

\caption{\textit{Left}: Cumulative fraction of \controlinj{} injections (i.e., ones without a $(3,3,0)$ mode in the ringdown) versus $(3,3,0)$ Bayes factor maximized over time. Since these injections have no $(3,3,0)$ mode in the ringdown, we expect the cumulative distribution of Bayes factors $\gtrsim 2$ when calculated after the merger to follow the black-dashed line. Shaded yellow regions show the 1-3$\sigma$ deviation regions. The dark blue markers/line show the cumulative distribution of Bayes factors for the \controlinj{} injections when maximized over time steps $\geq t_{\rm ref}^{\rm inj}$. For all injections $t \geq t_{\rm ref}^{\rm inj}$ was after the merger. This line follows the expected distribution. Light blue markers/lines show the distribution of Bayes factors when maximized over all times, including before merger. We get an elevated set of Bayes factors in this case, since the ringdown model is no longer valid before merger. The red vertical line shows the maximized Bayes factor for GW190521 (\eventbf{}), which occurred at $t_{\rm ref} + 6\,$ms. On average, we expect \expectedFA{} out of the 500 injections to have a Bayes factor greater than this; we find \measuredFA{} when maximized over $t_{\rm ref}^{\rm inj} \geq 0$. Based on this, we conclude that the quoted Bayes factor for GW190521 is statistically sound. \textit{Right}: Equivalent plot for \signalinj{} injections that have an $\ell, m = (3, 3)$ post-merger SNR of at least 4. Contrary to the \controlinj{} injections, we find the Bayes factors of this set of \numhiqinj{} injections to stay substantially above the noise background. When maximized over $t \geq t_{\rm ref}^{\rm inj}$, \measuredTA{} of the \numhiqinj{} injections have a Bayes factor greater than GW190521. This indicates that our pipeline is capable of detecting large Bayes factors when a $(3,3,0)$ mode is present.
}
\label{fig:cumbfs}
\end{figure*}

Figure~\ref{fig:cumbfs} (left) shows the cumulative fraction of \controlinj{} injections that yield Bayes factors larger than the value given on the x-axis. For larger $\bayesfac$, we expect the distribution to follow the line $1/x$. We show two results: one in which we maximize $\bayesfac$ over all times tested, $t-t_{\rm ref}^{\rm inj} \in [-9,24]\,$ms and one in which we maximize over times $t-t_{\rm ref}^{\rm inj} \in [0, 24]\,$ms. When maximizing over all times, we find that the injected distribution does not follow the expected distribution of $1/x$; more injections yield large Bayes factors than expected from noise. At the Bayes factor found for GW190521 (\eventbf{}), \measuredFAallt{} of the injections yield larger Bayes factor, whereas we expect $\sim 9$.\footnote{When maximizing over all time, we use 497 injections instead of 500. This is because one time point failed to converge for three of the injections. For all three injections, this time point was before $t_{\rm ref}^{\rm inj}$, which is why we are able to use all 500 injections when maximizing over $t \geq t_{\rm ref}^{\rm inj}$.} However, when maximizing over times $t-t_{\rm ref}^{\rm inj} \geq 0$, the injections show remarkable agreement with the expected distribution. Indeed, we find \measuredFA{} \controlinj{} injections yield a $\bayesfac \geq \eventbf$.

Maximizing over all grid times yields an excess of large Bayes factors because the negative times include times before merger for all of the injections (note the distribution of merger times for the NRSur7dq4 results in Fig.~\ref{fig:imrcompare-time_mass_ratio_final_mass}). As stated above, before merger the signal is not a superposition of QNMs. This breaks one of our assumptions above. Put another way -- anything not modeled by our signal model is ``noise''; in the pre-merger regime the ``noise'' is not Gaussian, and so larger Bayes factors can be obtained than otherwise expected.

However, this only happens if we sample \emph{before} the merger. By maximizing over $t-t_{\rm ref}^{\rm inj} \in [0, 24]\,$ms we are in the post-merger regime for 174 of the 500 the injections. In this case, we get good agreement with the expectations. We find similarly good agreement if we use our knowledge of the injections' coalescence time to only maximize over grid points that occur after $t_c^{\rm inj}$. Doing so introduces complications due to the fact that a different number of grid points is maximized over for each injection; see Appendix \ref{app:timemarg} for more details.

In order for the larger-than-expected false alarm rate to apply to GW190521, the time at which the maximum Bayes factor occurred (\eventbftime{}) would have to have been \emph{before} the merger. Of the 500 \controlinj{} injections only 15 had coalescence times after \eventbftime{}. We therefore conclude this scenario to be unlikely, and use the result when maximizing over $t_{\rm ref}^{\rm inj} \geq 0$. Given the excellent agreement between expectations and measurement, we conclude that our measured $(3,3,0)$ Bayes factor for GW190521 is valid.

To check that our code can recover large Bayes factors when a signal actually has a $(3, 3, 0)$ mode, we repeat this analysis on the \signalinj{} injections. The result is summarized in Fig.~\ref{fig:cumbfs} (right). As expected, the cumulative distribution of Bayes factors for \signalinj{} injections does not follow the distribution expected from noise. We also find there to be little difference between maximizing over $t-t_{\rm ref}^{\rm inj} \in [-9, 24]\,$ms and $t \geq t_{\rm ref}^{\rm inj}$. Of the \numhiqinj{} \signalinj{} injections, \measuredTA{} have Bayes factors larger than GW190521 when maximized over $t \geq t_{\rm ref}^{\rm inj}$ while \measuredTAallt{} have larger Bayes factors when maximized over all times.

\begin{figure*}[ht]
    \centering
    \includegraphics[width=\columnwidth]{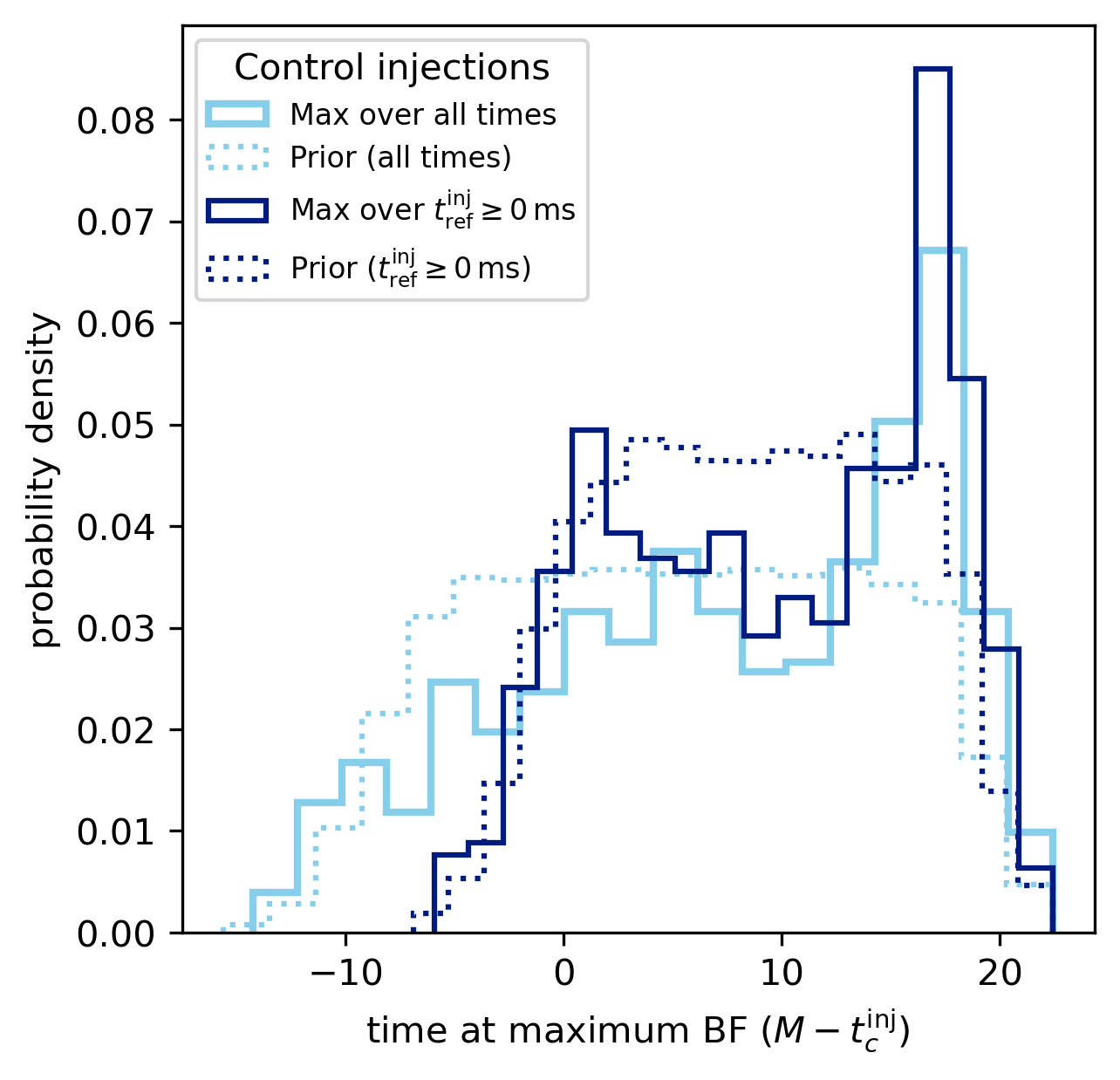} \hfill
    \includegraphics[width=\columnwidth]{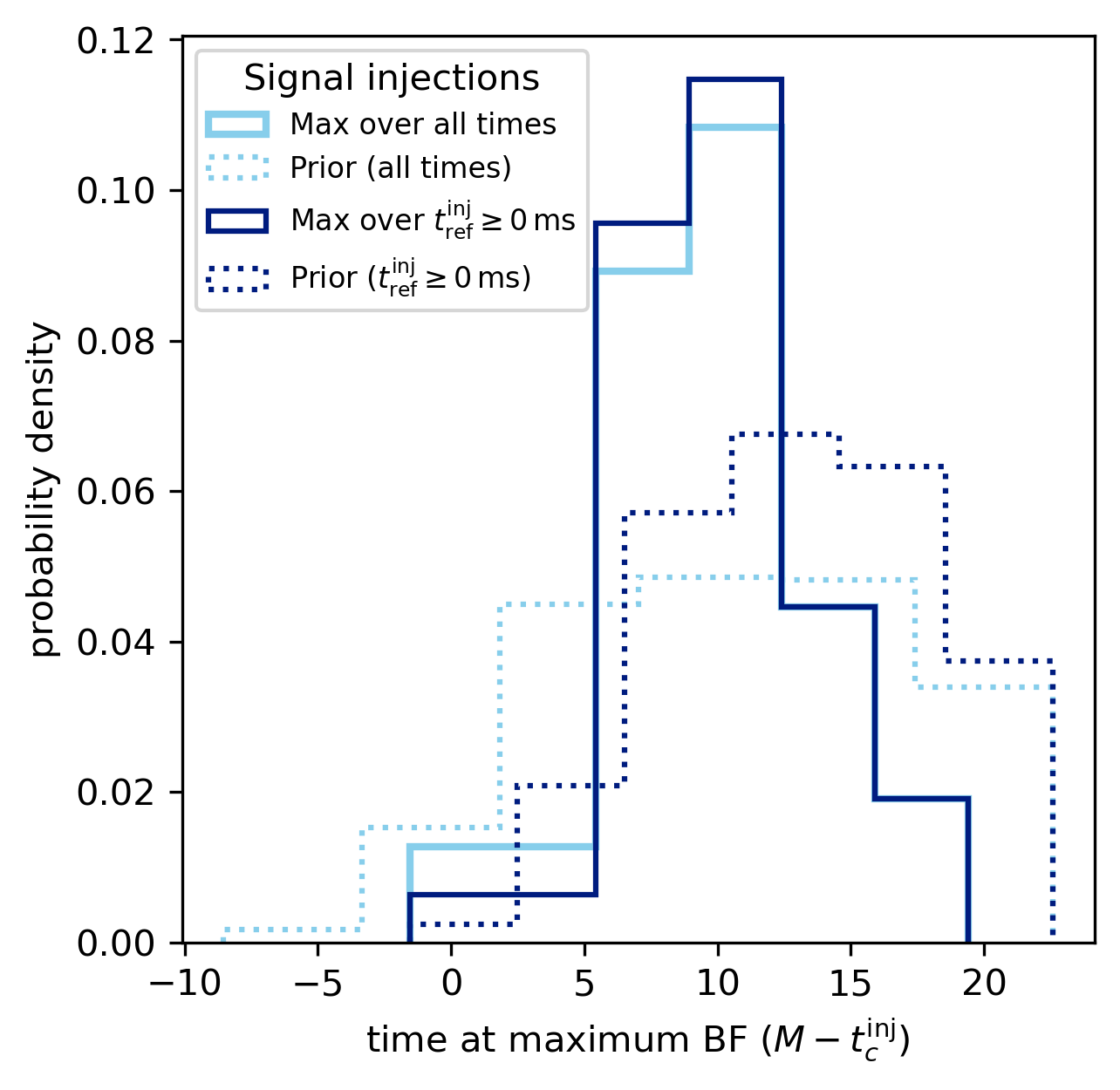}
    \caption{\textit{Left}: Distribution of times at which the maximum Bayes factor $Z_{220{+}330}/\max\{Z_{220}, Z_{220{+}221}\}$ occurred for the \controlinj{} injections. Times are quoted in units of mass $M$ after $t_c^{\rm inj}$. Shown are when the maximization is done over the entire time period sampled (light blue) and when it is done over just $t_{\rm ref}^{\rm inj}\geq 0$ (dark blue). The prior is also shown (dotted lines). The distribution is consistent with noise, as it largely uniform with peaks only near the boundaries. These peaks are due to the Bayes factors not being completely independent in time. \textit{Right}: Equivalent plot for the \signalinj{} injections. Here, the Bayes factors tend to peak at $\sim 10M$ after merger. This is consistent with expectations for when a QNM description of the post-merger signal that includes only fundamental modes becomes valid~\cite{Buonanno:2006ui,Berti:2007fi,Kamaretsos:2011um}.}
    \label{fig:maxbftime}
\end{figure*}

Figure~\ref{fig:maxbftime} shows the distribution of times at which the maximum Bayes factor occurs for the \controlinj{} (left) and \signalinj{} (right) injections. In the case of the \controlinj{} injections, no particular time is favored, as would be expected from noise. There is an excess in the distribution near the boundaries, but this is expected since the Bayes factor is not independent between consecutive time steps. If the Bayes factor is large at a given time, it is more likely to be elevated at surrounding times, due to the autocorrelation length of the template signal. In noise the maximum Bayes factor is therefore more likely to occur close to the boundary.

Conversely, for the \signalinj{} injections the distribution in time is peaked at $\sim 10M$ after merger. This is consistent with expectations. Numerical simulations have shown that a QNM description of the signal using only fundamental modes becomes valid between $10$ and $20M$ after merger~\cite{Buonanno:2006ui,Berti:2007fi,Kamaretsos:2011um}.
That the pipeline recovers large Bayes factors when the signal is present \emph{and} recovers those Bayes factors at $\sim 10M$ after merger validates its ability to recover the $(3,3,0)$ mode from a signal if it is present. As mentioned earlier, strictly speaking, this demonstrates the validity of our analysis when applied to the specific waveform models chosen, which are of course only approximations to exact general relativity; within these uncertainties, alternate explanations of GW190521 are not ruled out.  A detailed study quantifying such waveform uncertainties is beyond the scope of this work.

%\section{Parameter recovery}

\begin{figure*}
    \centering
    \includegraphics[width=\columnwidth]{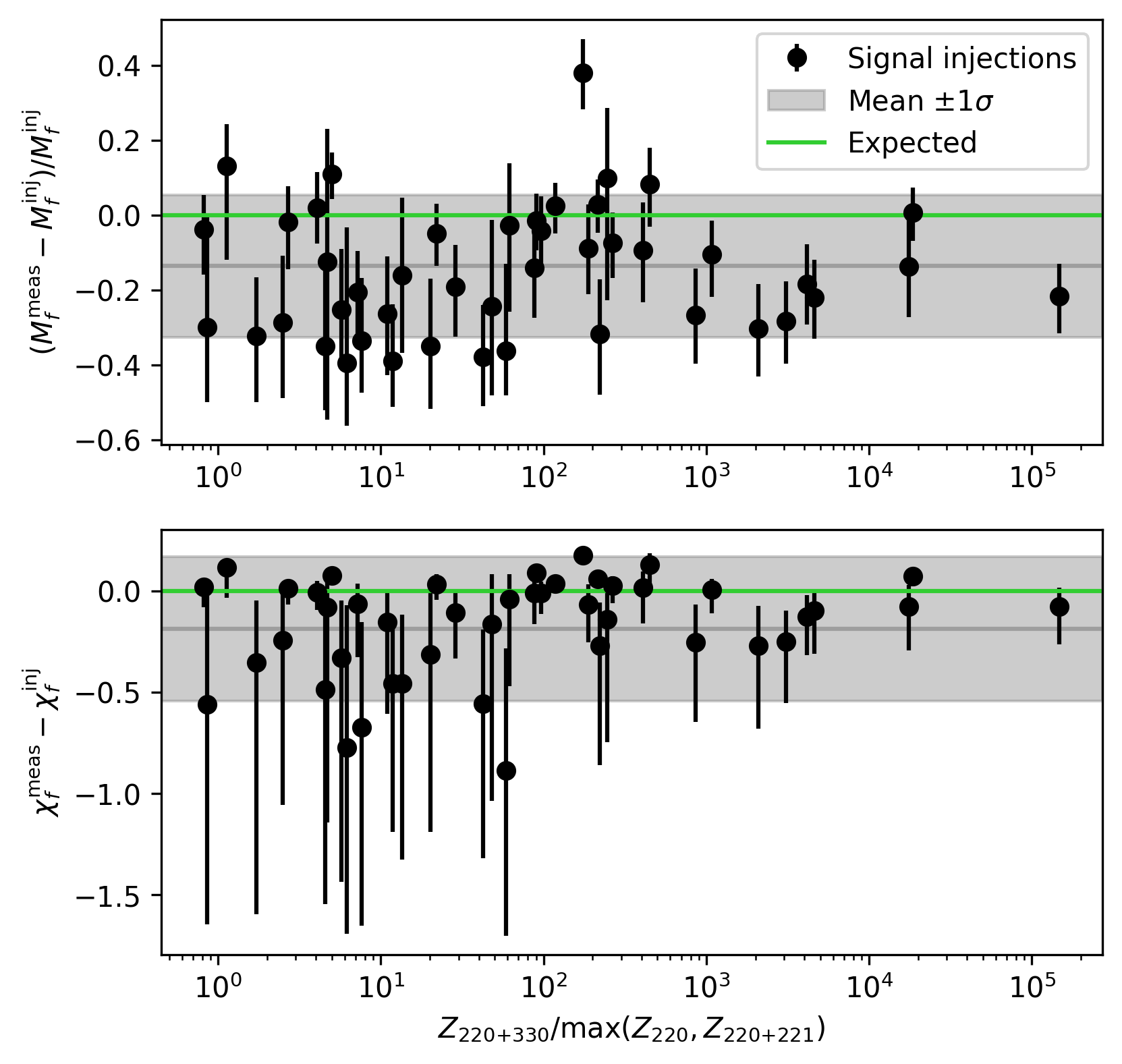} \hfill
    \includegraphics[width=\columnwidth]{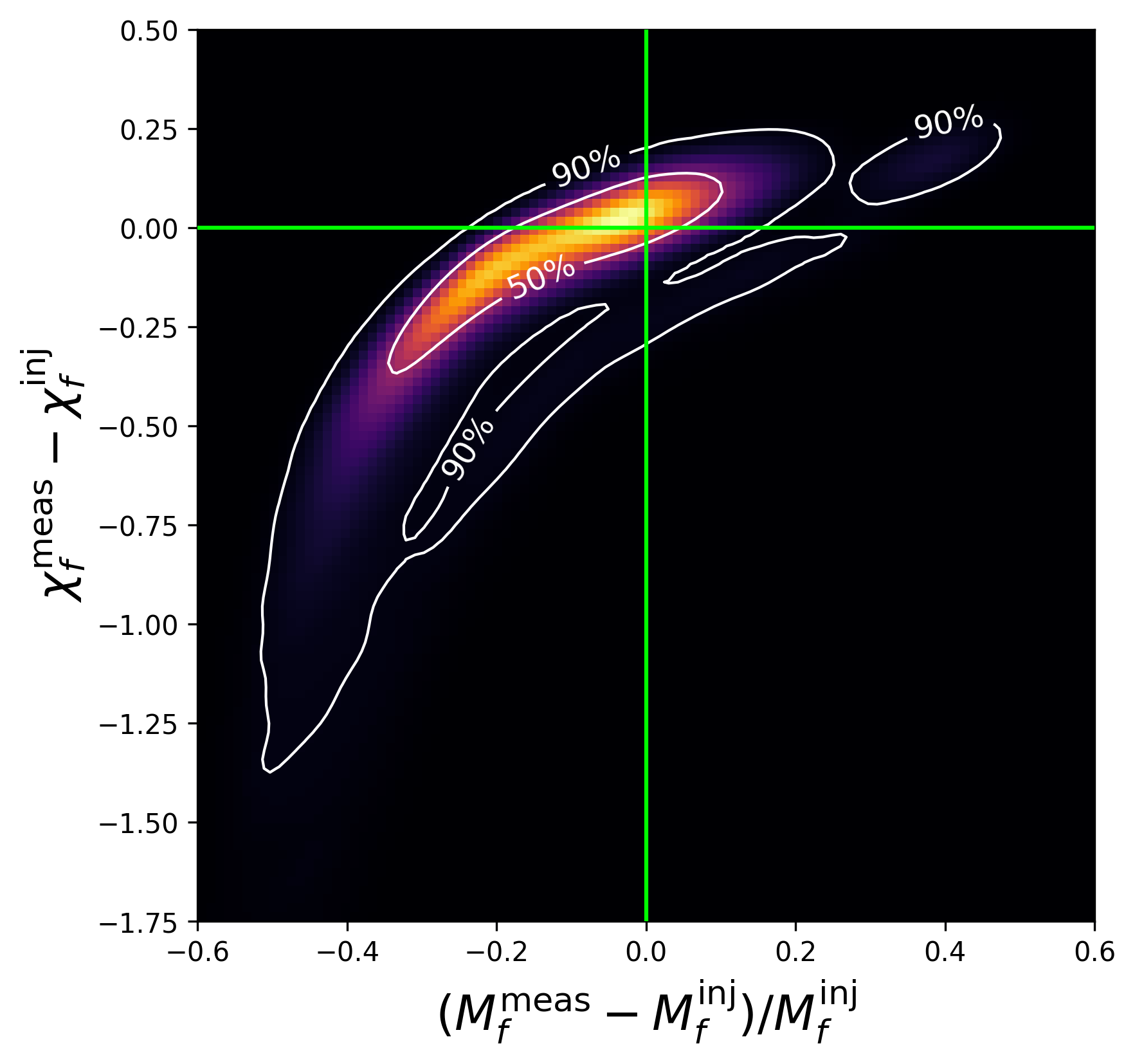}
    \caption{\textit{Left}: Marginalized fractional error in the recovered final mass $\delta M_f$ and marginalized absolute error in the final spin $\Delta \chi_f$ of each \signalinj{} injections using the $(2,2,0){+}(3,3,0)$, as a function of maximized Bayes factor $\max_t \mathcal{B}(330,t)$ [as defined by Eq.~\eqref{eqn:bayes_factor}]. The gray horizontal line and shaded region shows the mean $\delta M_f$ and $\Delta \chi_f$ $\pm 1\sigma$, averaged over posterior samples from all of the \signalinj{} injections. \textit{Right:} The 2D marginal distribution on both parameters, averaged over all of the \signalinj{} injections. This is obtained by adding together all of the injection's 2D marginal distributions on $\delta M_f$ and $\Delta \chi_f$. Zero error in both parameters (represented by the green lines) is within the 50\% credible region.}
    \label{fig:mass_spin_recovery}
\end{figure*}

Figure \ref{fig:mass_spin_recovery} illustrates the ability of the $(2,2,0){+}(3,3,0)$ model to recover the \signalinj{} injections' final mass $M_f$ and spin $\chi_f$ when the Bayes factor Eq.~\eqref{eqn:bayes_factor} is maximized over time. The figure shows 1D and 2D marginal distributions on the fractional error in recovered final mass $\delta M_f$ and the absolute error in final spin $\Delta \chi_f$. We find that, on average, the model reasonably recovers the mass and spin: zero is within 1$\sigma$ of the mean for both $\delta M_f$ and $\Delta \chi_f$ (left plot).  The $50\%$ credible region of the average 2D marginal posterior also contains zero (right plot).

\section{Tests of the no-hair theorem}
%*************************************************
\label{sec:testGR}

\begin{figure*}
    \centering
    \includegraphics[width=\columnwidth]{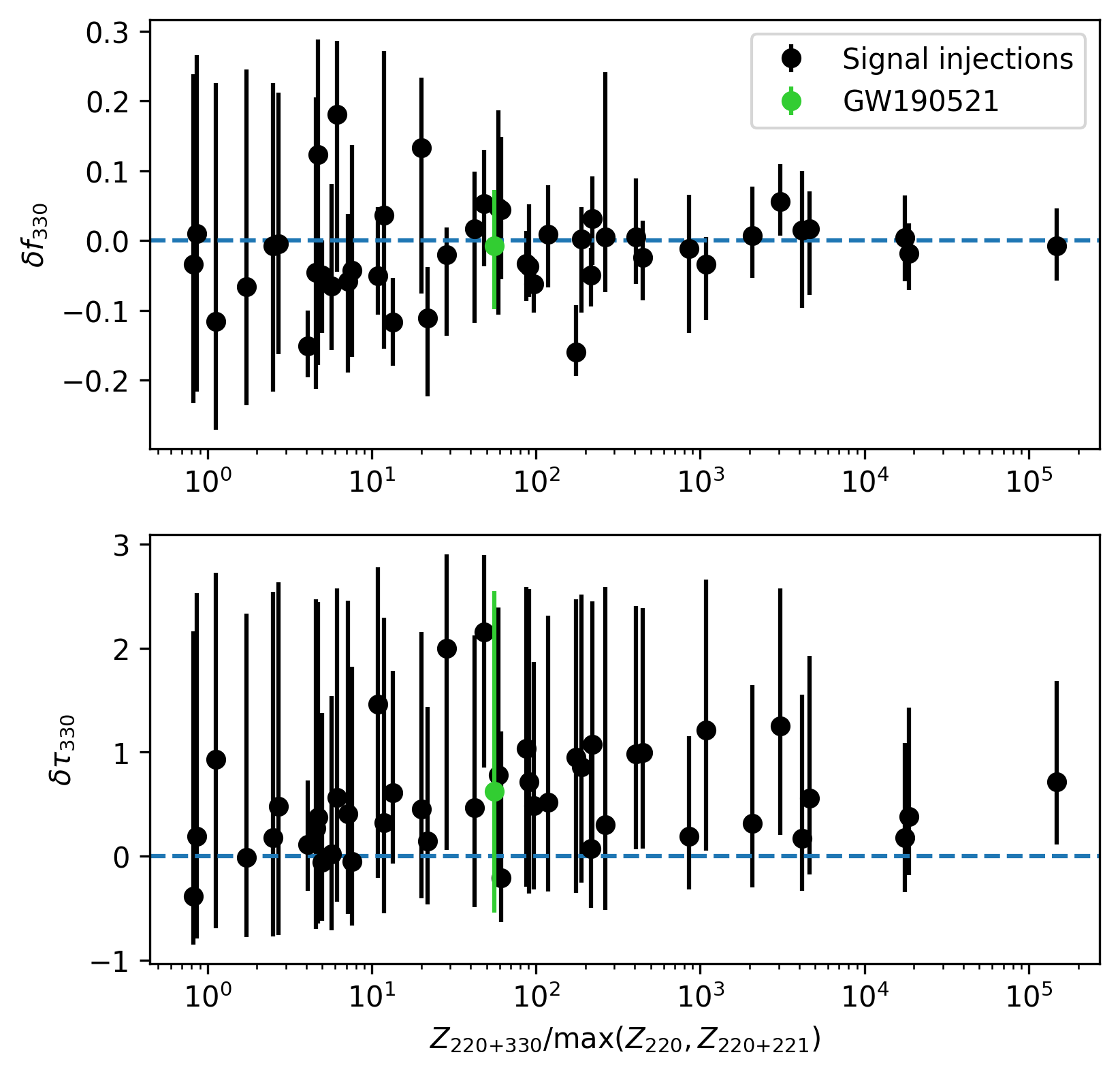} \hfill
    \includegraphics[width=\columnwidth]{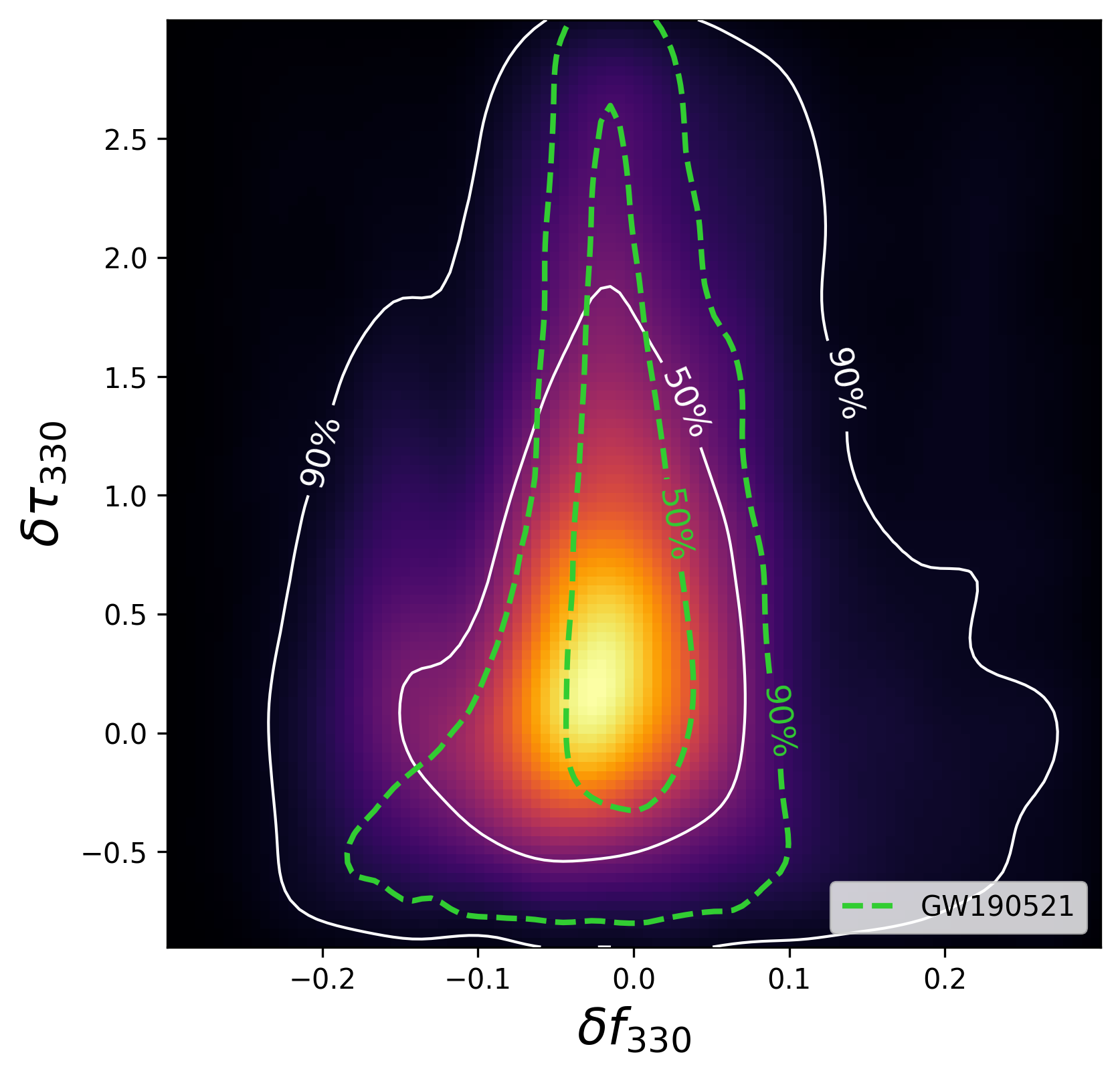}%
    \caption{Kerr deviation parameters $\delta f_{330}$ and $\delta \tau_{330}$ evaluated at the time of maximum $(2,2,0){+}(3,3,0)$ Bayes factor of the \signalinj{} injections. \textit{Left}: Marginalized distribution of the parameters as a function of Bayes factor for individual injections. Markers indicate the median value; errorbars show the $90\%$ credible range. For comparison, results from GW190521 are also plotted. \textit{Right}: The 2D marginal  distribution on $\delta f_{330}$ and $\delta \tau_{330}$ averaged over all the injections. This is obtained by adding together all of the injections' 2D marginal distributions. Dashed green contours show $50\%$ and $90\%$ credible region from GW190521. On average, $\delta f_{330}$ is concentrated close to zero (the expected Kerr value), and it approaches zero as the Bayes factor increases; damping time is not well measured. The constraints obtained on both parameters from GW190521 are consistent with the injections.}
    \label{figs:signal_no_hair_posteriors}
\end{figure*}

With more than one ringdown mode, a non-trivial test of the black hole no-hair theorem can be performed \cite{Dreyer:2003bv}.
Here we parameterize this test through the deviations $\delta f_\lmn$, $\delta \tau_\lmn$ associated with the measured frequency and damping time of the sub-dominant mode.

The deviation parameters are defined by $f_\lmn = (1 + \delta f_\lmn) f_\lmn(M_f, \chi_f)$ and equivalently for $\tau_\lmn$.
The mappings $f_\lmn(M_f, \chi_f)$ and $\tau_\lmn(M_f, \chi_f)$ to the black hole's final mass and spin $M_f, \chi_f$ assume Kerr and are calculated using the \texttt{pykerr} package~\cite{pykerr}. We fix the dominant mode's frequency and damping time to their Kerr values when doing this test, as has become common~\cite{LIGOScientific:2020ufj,Isi:2021iql}, while varying the final mass and spin. We then have four additional free parameters for each harmonic that is varied in the test, the deviation parameters $\delta f_\lmn, \delta \tau_\lmn$, and the mode's phase $\phi_\lmn$ and amplitude $A_\lmn$.

In Capano et al.\ we used GW190521 to apply this test to both the $(\ell, m, n) = (3,3,0)$ mode and the $(2, 2, 1)$ overtone of the dominant mode. Since these modes were best measured at different times, we did not test these modes simultaneously. Instead, the $(2,2,1)$ overtone deviation analysis was done when the Bayes factor for the overtone model was at a maximum, at $t_{\rm ref}-7\,$ms, while the $(3,3,0)$ deviation analysis was done at $t_{\rm ref}+6\,$ms. Since the $(2,2,0)+(2,2,1)$ and $(2,2,0)+(3,3,0)$ models were the most favored models at these respective times, we did not include any other modes when doing these tests. In other words, the intrinsic parameters for the $(3,3,0)$ test were $\{M_f, \chi_f, \delta f_{330}\, \delta \tau_{330}\, A_{220}, \phi_{220}, A_{330}/A_{220}, \phi_{330}\}$, and likewise for the $(2,2,1)$ test. See Table~\ref{table:prior} for the complete list of parameters and priors used.

The results of these tests on GW190521 yielded excellent constraints on $f_{330}$, with $\delta f_{330} = -1^{+8}_{-9}\,\%$ (90\% credible interval). The damping time was only weakly constrained, with $\delta \tau_{330}=60^{+190}_{-120}\,\%$. Previously, sub-$20\%$ constraints on any sub-dominant fundamental mode were not expected until at least the next generation of detectors~\cite{Berti:2016lat,Cabero:2019zyt}. This is because observations of binary black hole mergers prior to GW190521 had smaller total masses and were primarily equal mass ratio. The constraint on $f_{330}$ was also substantially better than has been obtained on the $(2, 2, 1)$ overtone with other events. The best single-event constraint on $\delta f_{221}$ as reported by the LVC was $\sim \pm 50\%$~\cite{LIGOScientific:2020tif}. Combining results over 21 observations yielded $\delta f_{221} = 1^{+27}_{-28}\%$~\cite{LIGOScientific:2021sio}, still a factor of $\sim3$ larger than the constraint on $\delta f_{330}$ obtained from GW190521 alone.

Here, we repeat our no-hair test on the \signalinj{} injections to check whether the constraints we obtained on the $(3,3,0)$ mode are reasonable at the signal strength of GW190521. Since the simulations satisfy general relativity (up to the modelling systematics of the waveforms we use) the results from these tests indicate what constraints can be obtained given the type of signal observed and the quality of the data around that time. For this test we use 44 of the \signalinj{} injections,\footnote{In total, there were 45 injections. We show results for 44 of these as the sampler was unable to converge for one of the injections.} generated with IMRPhenomTPHM including all its available modes. Each injection is analyzed at its time of maximum $(2,2,0)+(3,3,0)$ Bayes factor. This replicates what was done with GW190521 in Capano et al.

Results for $\delta f_{330}$ and $\delta \tau_{330}$ using the \signalinj{} injection set are summarized in Fig.~\ref{figs:signal_no_hair_posteriors}. Shown are 1D marginal results for each injection. We also show the 2D marginal distribution on these parameters, averaged over all the injections. Results from GW190521 are also plotted for comparison. On average, the injections' $(3,3,0)$ deviation parameters $\delta f_{330}$ and $\delta \tau_{330}$ are centered near zero, and $f_{330}$ becomes better constrained as the Bayes factor increases. The constraints derived from GW190521 are consistent with these results. This indicates that at the signal strength of GW190521, a constraint on $\delta f_{330} \sim 10\%$ is reasonable for GR signals even though only two QNM modes are included in the model at the time of the analysis.

Note that the $(2,2,0){+}(3,3,0)$ model is not a complete description of the signal either, even at $10\,M$ after merger. Other modes are present in these signals; as the SNR of signals increases it will be necessary to include additional modes in the model. However, our results indicate that at the signal strength of GW190521, the $(2,2,0){+}(3,3,0)$ model is sufficient to describe the signal and to constrain the deviations on the $(3,3,0)$ mode at the $\sim10\%$ level, as we obtained with GW190521.

\section{Conclusions}
%************************************************
\label{sec:conclusions}

The Bayes factor for two Kerr ringdown modes over just one mode was estimated to be $\eventbf$ for GW190521 in Capano et al.~\cite{Capano:2021etf} Here, we find that of 500 simulated signals with higher ringdown modes explicitly turned off, only \measuredFA{} are recovered with a Bayes factor higher than 56. Thus a statistic at least as significant as GW190521 occurs once in 50 times. We have in addition demonstrated that the analysis can detect a $(3,3,0)$ mode when it is present in the data.

We also developed a statistic $\zeta$ to quantify the agnostic analysis. This involves comparing the consistency of the two-dimensional frequency and damping-time posterior of the second mode with that predicted by the frequency and damping-time posterior from the first mode, assuming that they are the $(2,2,0)$ and $(3,3,0)$ modes of a Kerr black hole. We showed that this statistic was able to separate simulations with an observable $(3,3,0)$ mode from those without. Applying this to GW190521, we find that the probability of getting a $\zeta$ as large as that of GW190521 by chance from simulations that do not have a $(3,3,0)$ mode is $0.004$. 

Our results for the no-hair theorem test in Sec.~\ref{sec:testGR} show that the constraints on the $(3,3,0)$ mode observed for GW190521 in Capano et al. are consistent with what is obtained using simulated GR signals. It may be possible to use a QNM model at merger to constrain deviations from GR if enough overtones are included in the model. Determining the appropriate number of observable overtones without overfitting the data is a delicate question however, beyond the scope of this study. Our results here indicate that performing the no-hair test on fundamental modes, later in the signal, largely sidesteps such questions (at least at currently observable SNR), yielding much tighter and more reliable constraints.\footnote{Our results are broadly consistent with Ref.~\cite{JimenezForteza:2020cve}, even though they were studying non-spinning signals. There, it was found that the $(3,3,0)$ mode is the best observable non-dominant QNM if the mass ratio $\gtrsim1.2$ and the ringdown SNR $\gtrsim8$. Under these conditions they predicted that the $(3,3,0)$ frequency could be constrained to the $\sim10\%$ level. For context, GW190521's ringdown SNR was $\sim12$~\cite{Capano:2021etf}.}

Our simulation campaign supports the conclusion that a two mode model is observationally distinguishable from a single mode model in the ringdown of GW190521, and that the second mode is consistent with general relativity.

\section*{Acknowledgments}

We thank Gregorio Carullo for stimulating discussions at the beginning of this work. We also thank the Atlas Computational Cluster team at the Albert Einstein Institute in Hanover for assistance. C.C. acknowledges support from NSF award PHY-2309356. This research has made use of data obtained from the Gravitational Wave Open Science Center (https://www.gw-openscience.org/ ), a service of LIGO Laboratory, the LIGO Scientific Collaboration and the Virgo Collaboration. LIGO Laboratory and Advanced LIGO are funded by the United States National Science Foundation (NSF) who also gratefully acknowledge the Science and Technology Facilities Council (STFC) of the United Kingdom, the Max-Planck-Society (MPS), and the State of Niedersachsen/Germany for support of the construction of Advanced LIGO and construction and operation of the GEO600 detector. Additional support for Advanced LIGO was provided by the Australian Research Council. Virgo is funded, through the European Gravitational Observatory (EGO), by the French Centre National de Recherche Scientifique (CNRS), the Italian Istituto Nazionale di Fisica Nucleare (INFN) and the Dutch Nikhef, with contributions by institutions from Belgium, Germany, Greece, Hungary, Ireland, Japan, Monaco, Poland, Portugal, Spain.

\bibliography{reference.bib}
\appendix

\section{Effect of polarization marginalization on the Bayes factor}
%****************************************
\label{app:polmarg}

In the initial analysis in Capano et al.\ \cite{Capano:2021etf} we used \texttt{dynesty} to sample over all parameters for the Kerr analysis listed in Table~\ref{table:prior}. We found a maximum Bayes factor of $44^{+6}_{-5}$ in favor of the $(2,2,0)+(3,3,0)$ model at $t_{\rm ref}+7\,$ms.  However, this method proved time-consuming as the sampler struggled to converge for some mode combinations. The difficulty largely arises from the combination of the phases of the modes and the polarization angle. In particular, for GW190521 the phase of the dominant mode and the polarization are degenerate, as the polarization is not measured well due to the low SNR in the Virgo detector. This results in a banding pattern in the marginal likelihood between these parameters that is a challenge to sample.

Sampling over all parameters would have been unfeasible for the large number of injections we analyzed here. We therefore introduced a modified gating-and-in-painting model that numerically marginalized over the polarization using 1000 grid points. This marginalization technique was employed in the 3-OGC~\cite{Nitz:2021uxj} and 4-OGC analyses~\cite{Nitz:2021zwj}, where it was found to speed convergence for full IMR templates with sub-dominant modes. We are able to apply the same technique here because the dependence on the polarization is approximately constant over time for a short-duration event like GW190521, and so can be separated from the gating-and-in-painting procedure.

In implementing the polarization marginalization, we discovered that we obtained a larger Bayes factor for GW190521 one ms earlier, at $t_{\rm ref}+6\,$ms. To verify this, we repeated the $+6\,$ms and $+7\,$ms analysis 10 times using different starting seeds. We also repeated each analysis once with double the number of live points. We found consistently larger values at $+6\,$ms. Averaging the Bayes factors over the runs we obtained $56 \pm 1$ at $+6\,$ms and $45\pm1$ at $+7\,$ms, where the uncertainty is reported with 1$\sigma$. We further verified these Bayes factors by using the Savage-Dickey ratio on the $(3,3,0)$ amplitude posterior to estimate the Bayes factor, and obtained similar results as reported by \texttt{dynesty}'s estimate.

The result at $7\,$ms was consistent with our initial result in Capano et al., but the result at $6\,$ms was substantially higher. Our initial estimate for the Bayes factor at $+6\,$ms (without marginalization) was $40^{+5}_{-4}$. Evidently, without marginalization, the sampler had not fully converged at $6\,$ms, yielding an underestimate of the Bayes factor. Marginalization also affected our $(2,2,1)$ results: we found the Bayes factor for the $(2,2,1)$ mode peaked slightly earlier, at $t_{\rm ref}-7\,$ms instead of the $t_{\rm ref}-5\,$ms that we initially estimated.

Given the robustness of the new results under polarization marginalization, we quote the updated Bayes factor at $t_{\rm ref}+6\,$ms here for GW190521. We also updated Capano et al. to reflect these changes.\bigskip

\section{Maximizing the Kerr Bayes factor after merger}
%******************************************
\label{app:timemarg}

\begin{figure*}
    \centering
    \includegraphics[width=\columnwidth]{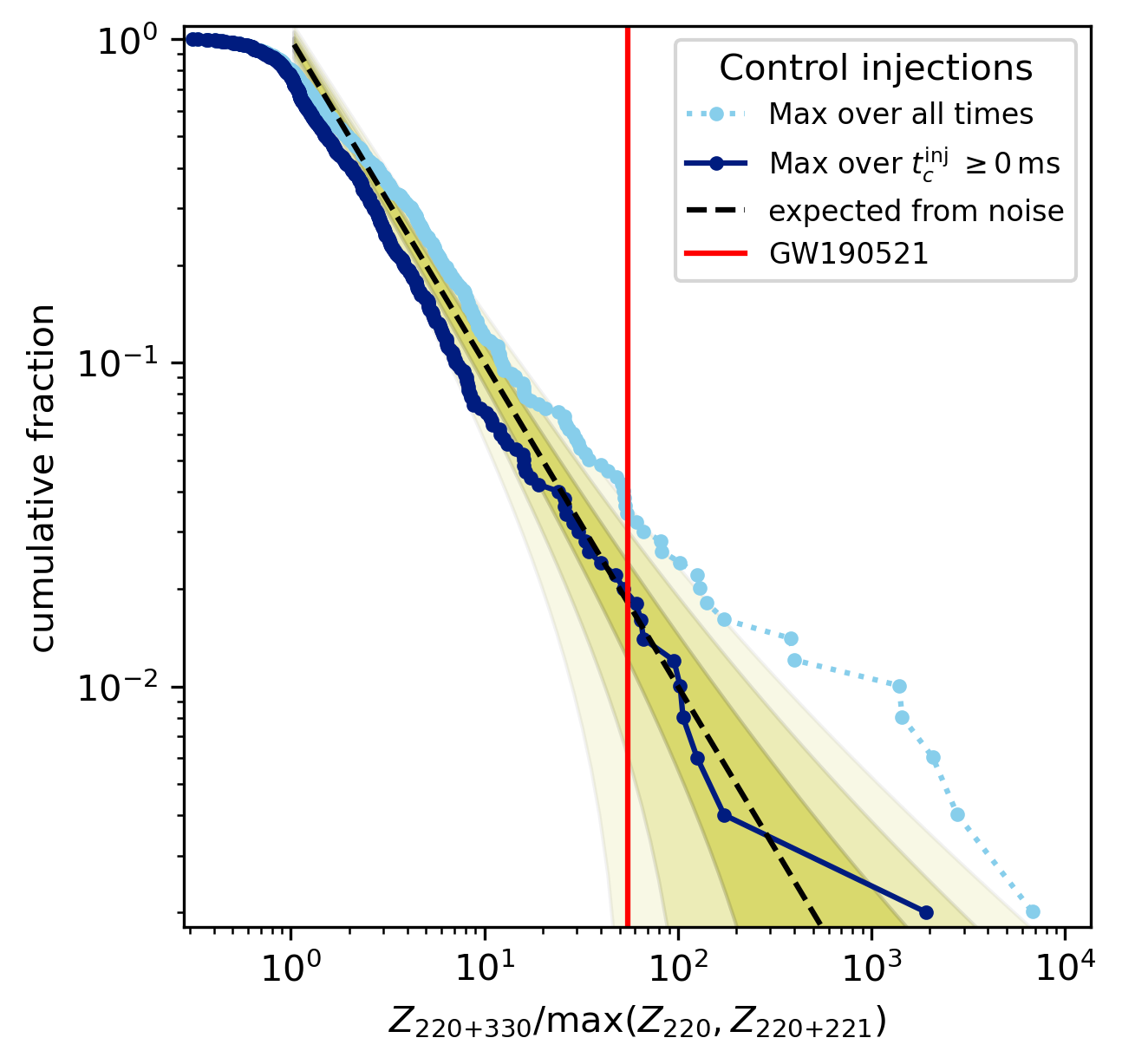} \hfill
    \includegraphics[width=\columnwidth]{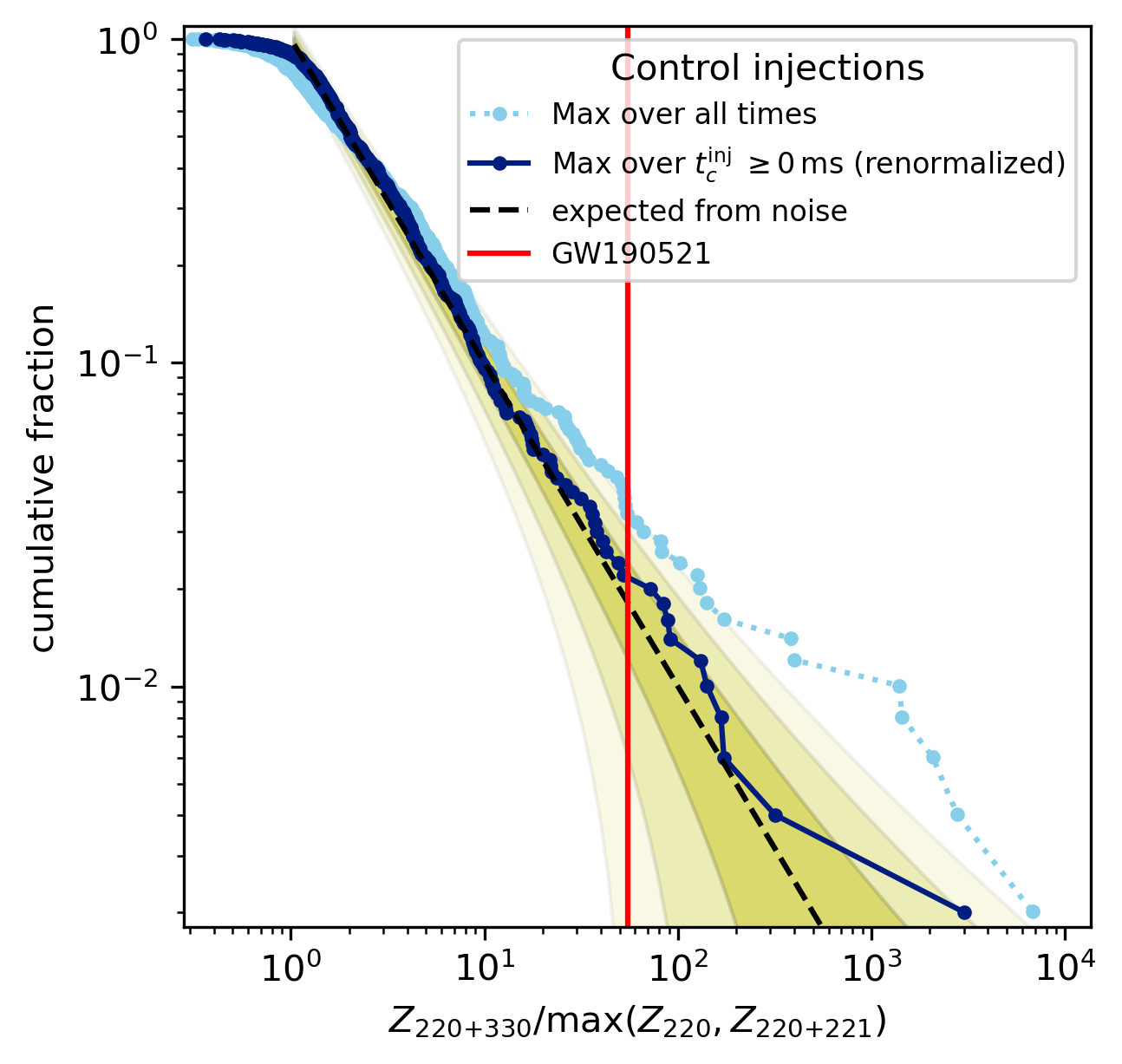}
    \caption{\textit{Left}: Same as Fig.~\ref{fig:cumbfs} (left), but with the maximization interval based on the injections' coalescence time $t_c^{\rm inj}$ (dark blue line). We find excellent agreement with the expected background distribution at large Bayes factors, but a $\sim 3\sigma$ downward deviation in the measured distribution at Bayes factors $\lesssim 20$. This deviation is due to the different maximization range for each injection.
    \textit{Right:} Renormalized version of the left plot. Here, we've accounted for variations in maximization interval across the \controlinj{} injections by multiplying their Bayes factor by $(\max N_{\rm grid})/N_{\rm grid}$, where $N_{\rm grid}$ is the number of time points maximized over. We find good agreement with the expected background at both large and small Bayes factors in this case.}
    \label{fig:cumbftc}
\end{figure*}

As discussed in Sec.~\ref{sec:statsig_Kerr}, we obtain good agreement between the expected distribution of Bayes factors and the measured distribution if we restrict the maximization interval to be strictly after the \controlinj{} injections' coalescence time $t_c^{\rm inj}$. The result is shown in the left plot of Fig.~\ref{fig:cumbftc}. Above Bayes factors of $\sim20$ we find excellent agreement with the background. Indeed, we find that 9  of the 500 injections have a Bayes factor larger than GW190521, exactly the amount expected by chance.

However, for Bayes factors $\lesssim 20$ there is a nearly $3\sigma$ downward deviation in the measured background. This deviation is due to the fact that differing numbers of grid points are maximized over when using the injection's coalescence time. For example, the maximization interval spans nine grid points (spanning $t_{\rm ref}^{\rm inj}+[0,24]\,$ms) for injections that have a $t_c^{\rm inj} \approx t_{\rm ref}^{\rm inj}$, whereas the interval is only two grid points for injections with $t_c^{\rm inj}\approx t_{\rm ref}^{\rm inj}+21$ ms. Although grid points are not independent of each other -- if a large Bayes factor exists at a particular point in time, there is a higher probability that its neighbors will also have larger Bayes factors -- they are not entirely dependent either. Due to the stochastic nature of the noise, there are random fluctuations in Bayes factors across time. Consequently, if a maximization interval covers fewer grid points, there are fewer opportunities to obtain larger Bayes factors.

Large Bayes factors are not strongly affected by differences in maximization interval, since there is a low probability that a noise fluctuation could produce a larger Bayes factor. This is evident in the left plot of Fig.~\ref{fig:cumbftc}. Conversely, smaller Bayes factors will be affected by this, hence the deviation at lower Bayes factors in that plot.

This issue can be corrected for by multiplying the Bayes factors of each injection by $(\max N_{\rm grid})/N_{\rm grid}$, where $N_{\rm grid}$ is the number of grid points maximized over for the given injection and $\max N_{\rm grid}$ is the largest number of grid points maximized over in the set. Renormalizing the Bayes factors yields the result shown in the right plot of Fig.~\ref{fig:cumbftc}. Now we find good agreement with the expected background and measured distribution at all Bayes factors. With this we find 10 \controlinj{} injections to have a larger Bayes factor when we expect 9.

Note that the normalization factor implicitly assumes that each grid point is independent of the others. As stated above, this is not the case. Since using this factor tends to overestimate the contribution, this is a conservative error.

Due to these complications we present in the main text the simpler maximization over $t_{\rm ref}^{\rm inj}\geq 0$.

\section{(2,2,0){+}(3,3,0)/(2,2,0)+(2,2,1) Bayes factors}
\label{app:330_221BF}

\begin{figure*}[ht]
\centering
\includegraphics[width=\columnwidth]{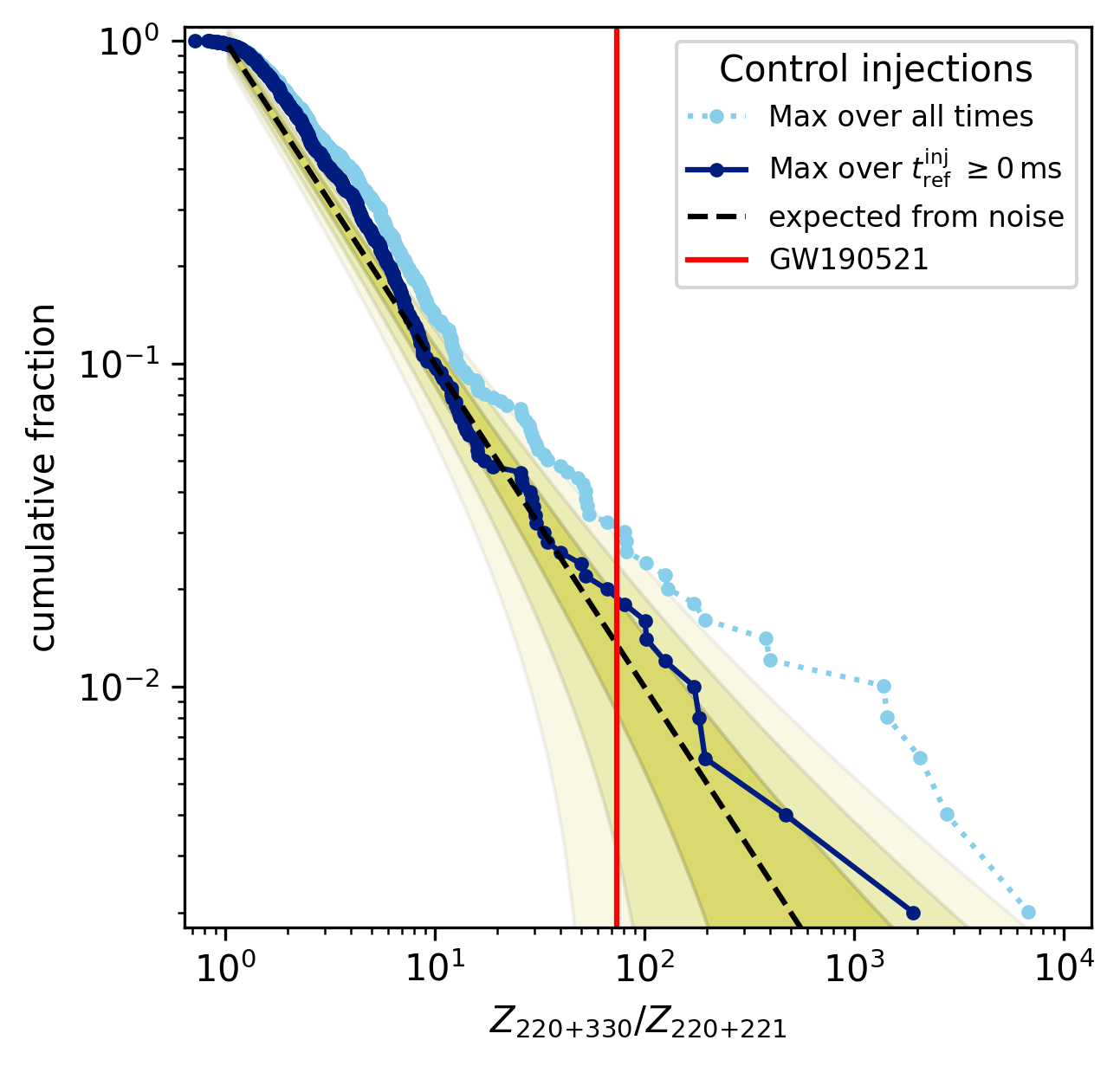} \hfill
\includegraphics[width=\columnwidth]{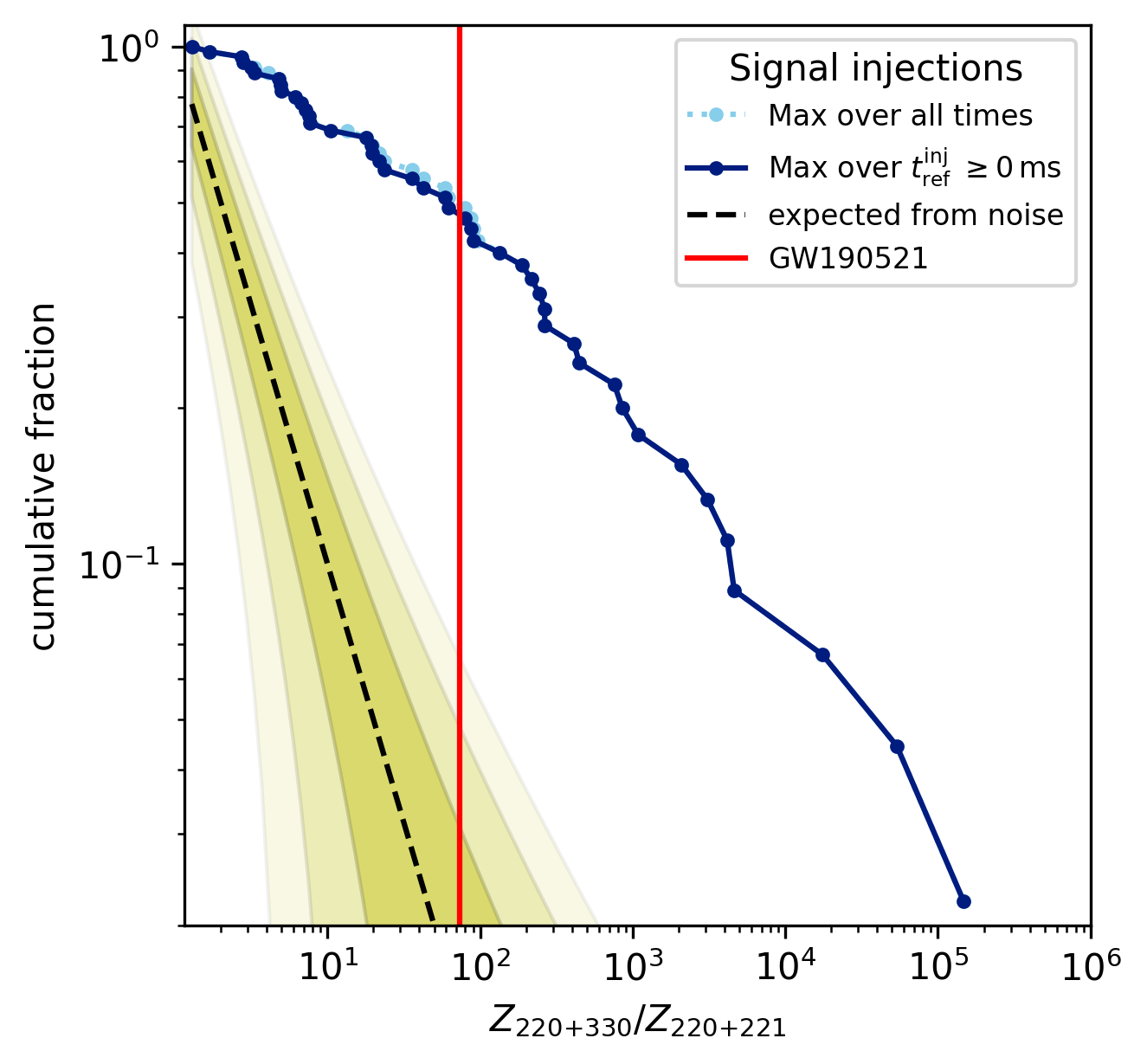}
\caption{Same as Fig.~\ref{fig:cumbfs} but for a Bayes factor comparing the $220{+}330$ model to just the $220{+}221$ model. The results are largely the same: the distribution of \controlinj{} injections matches the expected background whereas the distribution of \signalinj{} injections do not, as expected. The maximum Bayes factor for GW190521 (indicated by the vertical red line) is larger in this case, with a value of 74 instead of 56.}
\label{fig:cumbf330_221}
\end{figure*}

As stated in the main text, we compare the $(2,2,0)+(3,3,0)$ model to both the $(2,2,0)+(2,2,1)$ and $(2,2,0)$-only models in both this and our original GW190521 study~\cite{Capano:2021etf}, taking the minimum of the Bayes factor factors between them. This was chosen in order to evaluate the evidence for just a single additional fundamental mode being present, while also accounting for the $(2,2,0)$-only model's decreasing reliability close to merger. However, the Bayes factor that compares the evidence between $(2,2,0)+(3,3,0)$ model and the $(2,2,0)+(2,2,1)$ model, $Z_{220{+}330}/Z_{220{+}221}$, yields similar results as what we find for the minimum factor $Z_{220{+}330}/\max(Z_{220{+}221},Z_{220})$. This is illustrated in Fig.~\ref{fig:cumbf330_221}, which shows the cumulative fraction of \controlinj{} (left) and \signalinj{} (right) injections as a function of $Z_{220{+}330}/Z_{220{+}221}$.

As with $Z_{220{+}330}/\max(Z_{220{+}221},Z_{220})$, the distribution of $Z_{220{+}330}/Z_{220{+}221}$ for \controlinj{} injections follows the expected background when maximized over $t^{\rm inj}_{\rm ref}\geq 0\,$ms, while the \signalinj{} show and obvious deviation. The only notable difference is that the maximum Bayes factor for GW190521 (indicated by the vertical red line in the plots) is now 74 instead of 56 (although it still occurs at $t_{\rm ref}+6\,$ms). This further illustrates the conservative nature of the $Z_{220{+}330}/\max(Z_{220{+}221},Z_{220})$ statistic we used when analyzing GW190521.

\end{document}